%% file: Main.tex
\documentclass[sigconf]{aamas} 
\renewcommand\footnotetextcopyrightpermission[1]{} 
\usepackage{balance} 
\usepackage{color}
\input{preamble}

\title{Opponent-aware Role-based Learning in Team Competitive Markov Games}

\author{Paramita Koley}
\affiliation{
   	\institution{IIT Kharagpur}
  \city{Kharagpur}
  \country{India}}
\email{paramita.koley@iitkgp.ac.in}

\author{Aurghya Maiti}
\affiliation{
  \institution{Columbia University}
  \city{New York}
  \country{USA}}
\email{am5887@columbia.edu}

\author{Niloy Ganguly}
 \affiliation{
 	\institution{IIT Kharagpur}
 	\city{Kharagpur}
 	\country{India}}
 \email{niloy@cse.iitkgp.ac.in}
 
\author{Sourangshu Bhattacharya}
 \affiliation{
 	\institution{IIT Kharagpur}
	\city{Kharagpur}
	\country{India}}
 \email{sourangshu@cse.iitkgp.ac.in}

\begin{abstract}
    Team competition in multi-agent Markov games is an increasingly important setting for multi-agent reinforcement learning, due to it’s general applicability in modeling many real-life situations. Multi-agent actor-critic methods are the most suitable class of techniques for learning optimal policies in the team competition setting, due to their flexibility in learning agent-specific critic functions, which can also learn from other agents. In many real-world team competitive scenarios, the roles of the agents naturally emerge, in order to aid in coordination and collaboration within members of the teams. However, existing methods for learning emergent roles rely heavily on the Q-learning setup which does not allow learning of agent-specific Q-functions. In this paper, we propose RAC, a novel technique for learning the emergent roles of agents within a team that are diverse and dynamic. In the proposed method, agents also benefit from predicting the roles of the agents in the opponent team. RAC uses the actor-critic framework with role encoder and opponent role predictors for learning an optimal policy. Experimentation using 2 games demonstrates that the policies learned by RAC achieve higher rewards than those learned using state-of-the-art baselines. Moreover, experiments suggest that the agents in a team learn diverse and opponent-aware policies.
\end{abstract}

\keywords{MARL, Team competition, Emergent roles, Actor-critic}

\newcommand{\BibTeX}{\rm B\kern-.05em{\sc i\kern-.025em b}\kern-.08em\TeX}

\begin{document}


\fancyhead{}


\maketitle 

\input{000intro}

\input{001related_work}
\newcommand{\cC}{\mathcal{C}}
\newcommand{\bx}{\mathbf{x}}
\newcommand{\ba}{\mathbf{a}}
\newcommand{\br}{\mathbf{r}}
\newcommand{\cD}{\mathcal{D}}
\newcommand{\by}{\mathbf{y}}

\section{Opponent-aware Role-based MARL}
\label{sec:problem_formulation}

In this section,  we develop the proposed algorithm \our\ for Multi-agent Policy learning with emergent roles in the team-competitive Markov games setting \cite{lowe2017multi}.

\input{100prelim}

\input{101maac}

\input{102roma}

\input{103rac}
\input{200exp}

\input{300conclusion}

\bibliographystyle{ACM-Reference-Format} 
\bibliography{Main}


\end{document}

%% file: 000intro.tex
\section{Introduction}

Many real-world scenarios can be modeled as \textbf{team-competitive Markov games}~\citep{lowe2017multi} (also called \textit{mixed cooperative-competitive games}), where each agent receives an individual reward for its actions and a shared reward with other agents in the same team for a joint outcome. For example, while playing soccer, each player may receive a reward for their individual performance and a shared reward for the team's victory. In some social interactions, e.g., companies vying for customers, each employee may receive a reward for their individual performance and a shared reward for the team performance. 
\textbf{Roles} emerge naturally in such reward scenarios, since  they allow agents within a team to cooperate effectively. For soccer players, the roles could be ``striker'' and ``defender'', while for company employees, the roles could be ``production'' and ``sales''. In this work, we study the problem of learning policies with emergent roles in a reinforcement learning (RL) setting for team-competitive Markov games.


We use multi-agent reinforcement learning (MARL) to learn the optimal policies of individual agents in this setting. While many popular deep MARL techniques exist for the shared-reward setting
~\citep{gupta2017cooperative, mordatch2018emergence, rashid2018qmix,son2019qtran, wang2019learning}, relatively fewer techniques exist for the mixed-reward setting.
%
Multi-agent actor-critic algorithms~\citep{lowe2017multi,iqbal2019actor}, following the centralized training and decentralized executions concept, are the most popular MARL approaches in the mixed-reward setting, which is needed to model the team-competitive scenarios. MADDPG \citep{lowe2017multi}
is one of the earliest techniques for learning deterministic policies in mixed reward settings using the actor-critic framework. MAAC ~\citep{iqbal2019actor}
introduces attention mechanisms in critic modules to selectively pay attention to information collected from other agents, resulting in better scalability in complex multi-agent environments.
However, to the best of our knowledge, role-oriented policy learning in the actor-critic framework, particularly in competitive games, is yet unexplored.

On the other hand, the concept of predefined roles for agents has been explored to reduce design complexity in many MARL systems~\citep{spanoudakis2010using, bonjean2014adelfe}.
There are few works addressing dynamic role emergence based MARL in domain-specific problems~\citep{barrett2015cooperating,leottau2015study,urieli2011optimizing,roy2020promoting,zhang2021hierarchical,khaleefah2021exploring}.
Recently, Wang et al.~\citep{wang2020roma}  introduce automatic role emergence in MARL for cooperative tasks in their system called \roma. It proposes a latent role-based task decomposition among the agents built over the QMIX~\citep{rashid2018qmix} platform.  
In a similar line, Wang et al.~\cite{wang2020rode} propose RODE, a bi-level role-based learning framework where joint action-space is decomposed into restricted role-action spaces by clustering actions by their effects. Finally, it employ a role-selector to search role at a smaller role space, while role policies learn at a reduced action-observation space, significantly improving learning efficiency. Liu et al.~\citep{liu2022rogc} present ROGC, a role-oriented graph convolution based MARL framework, where roles are generated by classifying agents, graph convolution module is emplyed for efficient intra-role communication and finally effective individual policies are generated by embedding role information into the algorithm.
However, all of the above frameworks are built over QMIX, where a global utility function is composed of per-agent local utility functions, and the composition relies upon the assumption that all agents share the same reward. Hence, this setting can not be immediately extended to a mixed-reward setting, prohibiting role-oriented policy learning in the presence of opponents through joint learning of any form.
Recently, opponent-aware actor-critic algorithms have been explored in many settings, including 
learning time-dynamical opponent models TDOM-AC~\cite{wen2019probabilistic} , maximum entropy regularized opponent models ROMMEO~\cite{tian2019regularized}, opponent model employing variational Bayes PRR~\cite{tian2022multi}, etc. These works validate the necessity of opponent models for improving performance in competitive games. However, they are primarily limited to one-to-one competitions and do not consider role-aware opponent modeling. However, in a team competition setting, where each agent employs a role-aware policy within the team, considering roles in opponent modeling can benefit agents by reducing design complexity.

In this work, we propose \our, an actor-critic algorithm that combines role learning and role-aware opponent modeling in mixed-reward setting. 
\our\ uses the actor-critic framework based on MAAC \cite{iqbal2019actor}, which allows the critic to use joint observations of all agents, hence incorporating the opponent's experience in the critic. 
Our primary contributions are two-fold. First, we incorporate role-encoder in the actor-critic framework that learns dynamic and diverse emergent roles per team. 
Moreover, for opponent modeling, we incorporate an opponent role predictor network for each team, which is accessible to the policy at the execution time. 
The opponent role prediction network is trained jointly along with the role encoder, critic, and policy network. In particular, during centralized training, the opponent role predictor is trained to mimic the actual role encoding distribution of the opponent. During execution, that benefits agents to select their response.  
The learned roles are dynamic depending on the agent's trajectory and diverse between agents within the team. Experiments using the proposed episodic board games, \ourgame\, and \market, show that \our\ agents outperform \bsln\ agents in direct competitions. Furthermore, our experiments demonstrate that the dynamic roles learned by \our\ conform to the intuitive role-specific behaviors for each game, thus confirming the effectiveness of emergent roles.

In summary, our contributions are as follows:
\begin{itemize} 
\item  We propose \our{,} the first  opponent-aware role-oriented policy learning in actor-critic framework, designed for mixed reward setting, as per our knowledge. 
\item \our\ encourages diverse roles within a team and ensures that the role encoding captures the agent trajectory.
\item Most importantly, \our\ learns an opponent role predictor module, utilizing opponent role encodings during the centralized training that it employs in policy during the execution time. Eventually, the opponent role predictor guides agents in choosing strategic responses against adversaries.
\item Empirically, we validate improved policy learning of \our{, } through increased reward as well as demonstrate the identifiability of learned roles with intuitive role-specific behaviors of agents in two episodic board games, \ourgame\ and \market{.} 
\end{itemize}

%% file: 001related_work.tex
\section{Related Literature}
Multi-agent Actor-critic methods, primarily following the concept of centralized training and decentralized executions, are one of the most dominant and popular techniques in deep Multi-agent Reinforcement Learning (MARL) approaches~\citep{sukhbaatar2016learning,gupta2017cooperative,lowe2017multi,foerster2017stabilising,mordatch2018emergence,rashid2018qmix,foerster2018counterfactual}.  
Among these works, MADDPG~\citep{lowe2017multi}, a multi-agent adaptation of DDPG~\citep{lillicrap2016continuous}, considers continuous action space and deterministic policies in mixed-reward settings and successfully trains the agents complex coordination strategies in cooperative as well as competitive scenarios. 
MAAC~\citep{iqbal2019actor}, an extension of SAC~\citep{haarnoja2018soft}, introduces attention mechanism in critic modules to selectively pay attention to information collected from other agents, resulting in massive scalability in complex multi-agent environments. Also, there exists a series of value-function-factorization-based methods that train decentralized policies in a centralized end-to-end fashion, employing a joint action value network~\citep{rashid2018qmix}.
There are many follow-ups on actor-critic-based MARL algorithms, addressing a variety of issues, namely SAC~\citep{qu2020scalable} for improving scalability, SEAC~\citep{christianos2020shared} for sharing experience, LICA~\citep{zhou2020learning} for the credit assignment problem, TESSERACT~\citep{mahajan2021tesseract} for tensorizing the critics, Bilevel Actor Critic~ \cite{zhang2020bi}  for multi-agent coordination problem with unequal agents,
DAC-TD~\cite{figura2022cooperative} for training agents in a privacy-aware framework, VDAC~\cite{su2021value} for combining value-decomposition framework with actor-critic, Scalable Actor Critic~\cite{lin2021multi} for scalable learning in stochastic network of agents, etc.
However, none of these works consider the role emergence paradigm in their model. 

On the other hand, role division is quite common and efficient for accomplishing any complex collective task in human society~\citep{butler2012condensed}. Following that, many multi-agent systems also decompose the task into pre-defined roles, resulting in reduced design complexity and faster learning ~\citep{wooldridge2000gaia,padgham2002prometheus,cossentino2005passi,spanoudakis2010using,bonjean2014adelfe}. 
In recent times, Majumdar et al.~\citep{majumdar2020evolutionary} present a line of work where MADDPG agents learn different strategies by de-coupling and automatically weighting individual and global goals in a population-based training paradigm. 
Liu et al.~\citep{liu2021coach} consider the problem of coordination with dynamic composition, where a coach agent with global view distribute individual strategies to dynamically varying player agents. Jiang et al. ~\citep{jiang2021emergence} promote sub-task specialization via emergence of individuality through generating dynamic intrinsic rewards. Su et al.~\cite{su2022divergence} propose another actor-critic algorithm where agents learn diverse policies in cooperative team games.
Also, there is a series of works for role-oriented MARL for specialized domains like Robo-soccer ~\citep{barrett2015cooperating,leottau2015study,urieli2011optimizing,ossmy2018variety}, football environment~\citep{roy2020promoting}, swarn systems~\citep{zhang2021hierarchical}, image feature extraction tasks~\citep{khaleefah2021exploring} showing how complex policies can be learned by decomposing them into simpler sub-policies.

 In contrast, Wang et al. propose \roma~\citep{wang2020roma} and RODE~\citep{wang2020rode}, two dynamic and adaptive role-oriented MARL frameworks built over QMIX~\citep{rashid2018qmix}  in a quite general setting and show dynamic role emergence within the team on SMAC benchmark tasks. 
Recently, Liu et al. present ROGC~\citep{liu2022rogc}, a graph convolution based role-oriented MARL framework. However, all of them are inherently restricted to shared reward settings, i.e., only cooperative tasks.

Our work closely resembles~\citep{wang2020roma}, with the main distinction being that our work incorporates opponent modeling along with adopting team-based role learning to the actor-critic framework to exploit the advantage of training in the presence of opponents.
There are few recent works exploring opponent modeling in actor-critic framework, namely, TDOM-AC~\cite{tian2022multi} for learning time dynamical opponent model,  ROMMEO~\cite{tian2019regularized} for learning maximum entropy regularized opponent model, PRR~\cite{wen2019probabilistic} for learning opponent model employing variational Bayes, to name a few. They primarily consider the setting of two-player competitive games, where each agent maintains a prototype of opponent policy. However, they do not consider team competition settings. Extending their framework to our role-aware competitive team setting is not immediate. 

%% file: 100prelim.tex
\subsection{Preliminaries}

\begin{table}[t]
	\begin{tabular}{c l}
		\toprule
		Symbol & Meaning \\\midrule
		$\Scal,\Acal$ & State and action Space \\
		$\Pcal,\Rcal$ & Transition and reward functions \\
		$N, K$ & Total Number of agents and teams \\
        $\Ncal(k)$ & Set of agents in team $k$\\
		$\obar=(o_1,\dots,o_N)\in \Ocal $ & observation vector\\
		$\abar=(a_1,\dots,a_N)\in \Acal $ & action vector\\
		$\pi=(\pi_1,\dots,\pi_N)$ & Agent policies  \\
		$\tau^t=(\tau_1^t,\dots,\tau_N^t)$ & Trajectory till time $t$\\
		$Q=(Q_1,\dots,Q_N)$ & Value-action function \\
		$h_{S}=(h_S^1,\dots,h_S^K)$ & Self-role encoder \\
		$h_{O}=(h_O^1,\dots,h_O^K)$ & Opponent-role encoder \\
		$\rho=(\rho_1,\dots,\rho_N)$ & Self-role \\
        $\hrho=(\hrho_1^o,\dots,\hrho_N^o)$ & Predicted opponent-role \\
		$\lambda$ & Decay factor \\
		$\Lcal_{Q}$ & Critic loss \\
		$\Lcal_{MI}$ & Mutual information loss\\
		$\Lcal_{D}$ & Divergence loss \\
		$\Lcal_{Opp}$ & Opponent loss \\
		\bottomrule
	\end{tabular}
	\caption{List of important notations.}
	\label{tab:notations}
\end{table}

A Markov game is characterized by the tuple $\langle \Scal,\Acal,\Pcal,\Rcal \rangle$, where $\Scal$ denotes the set of states, $\Acal$ denotes the set of actions for each of the $N$ agents.  Hence the joint space of actions becomes $\Acal^N$.
The state transition function maps every state and joint-action combination to a probability over future states. $\Pcal \colon \Scal \times \Acal^N \rightarrow \PP(\Scal)$. 
In case of team-competitive games (also called mixed cooperative-competitive
games \cite{lowe2017multi}), let $\Kcal$ denote the set of $K$ teams, and $\Ncal(k)$ be the set of agents of team $k$,$k \in \Kcal$. 
The reward function  $\Rcal = (r_1,\dots,r_N) \colon \Scal \times \Acal^N  \rightarrow \RR^N $ specifies the reward scheme ($r_i$) for each agent $i$.
The per-agent reward function allows modeling of team competitive games~\citep{lowe2017multi}, which are the focus of this paper.
Also, in the decentralized execution setting, let $o_i \in \Ocal$ denote the observation for agent $i$. 
Each agent $i$ learns a policy  $\pi_i \colon \Ocal \rightarrow \PP( \Acal )$, which is a mapping  from it's own observation to a distribution over it's action set $\Acal$.  
We also define the trajectory $\tau_i(t) \in T = (\Ocal\times\Acal)^*$ for each agent $i$ till time $t$ as a sequence of observation-action pairs $\{ (o_i^j,a_i^j),  j=1,\dots,t \}$. We also denote the joint action and observations vectors as: $\bar{a}=(a_1,\dots,a_N)$ and $\bar{o}=(o_1,\dots,o_N)$. Agents aim to learn their policy $\pi_i$ by optimizing the agent-specific cumulative discounted reward function $\Jcal_i$: 
\begin{align}
  \Jcal_i(\pi_{i}) &= E_{\abar \sim \pi, s \sim \Pcal} \left[ \sum_{u=0}^{\infty} \gamma^{u} r_{i} (s^u, \abar^u) \right]
\label{eq:J_pol_grad}
\end{align}

\sloppy

%% file: 101maac.tex
\subsection{Multi-Agent Actor Critic}

Actor-critic methods and their extensions are widely used in multi-agent reinforcement learning \cite{haarnoja2018soft,iqbal2019actor} due to their flexibility and low variance reward estimates, which lead to faster learning of policy parameters. A popular extension is the soft actor-critic \cite{haarnoja2018soft}(SAC), where the reward of policy gradient (Eqn \ref{eq:J_pol_grad})  is augmented with the expected entropy of the policy distribution in each state. Hence their reward function becomes:
\begin{align}
  \Jcal_i(\pi_{i}) &= E_{\abar \sim \pi, s \sim \Pcal} \left[ \sum_{u=0}^{\infty} \gamma^{u} (r_{i} (s^u, \abar^u) + \alpha \mathcal{H}(\pi_i(s^u)) \right]
\end{align}
where $\alpha$ is the user defined regularization coefficient. Note that they do not explicitly discuss multi-agent RL in team-competitive setting.
 Subsequently, multi-attention actor critic (MAAC) \cite{iqbal2019actor} extended SAC to support centralized training of agents' policies with individual rewards. The learning algorithm in MAAC has two broad components: (1) SAC-style update of policy $\pi_i$ for the $i^{th}$ agent and (2) Attentive learning of the critic function $Q_i$ for the $i^{th}$ agent. The policy update is carried out using the policy gradient given by:
\begin{multline}
  \nabla_{\pi_i} \Jcal_i(\pi_{i}) = E_{\obar \sim D,\abar \sim \pi} [\nabla_{\pi_i} \log (\pi_{i} (a_{i}|o_{i})
  ( Q_i(\bar{o},\bar{a}) \\
   - b(\obar,\abar_{-i}) 
   - \alpha \log (\pi_i (a_i | o_i ))) ]
\label{eq:pol_grad_maac}
\end{multline}
where $\alpha$ is the regularization constant for the entropy term. $D$ is the replay buffer that contains the tuple $(\obar, \abar, \rbar, \obar^{'})$ for each step of the episodes generated by the policies till the current RL training time $t$. Here action $\abar$ at observation $\obar$ generates reward of all agents $\rbar$ and next observation $\obar^{'}$. $b(o,a_{-i})$ is the multi-agent baseline calculated as follows. 
\begin{align} 
    b(\obar,\abar_{-i}) &= E_{a_i \sim \pi_i(o_i) } \left[ Q_i(\obar,(a_i,\abar_{-i}))\right] \\
    &= \sum_{a_i \in \mathcal{A}} \pi_i(a_i |  o_i) Q_i(\obar,(a_i,\abar_{-i}))
\end{align}
The baseline is used calculate the advantage function, which helps in solving the multi-agent credit assignment problem \cite{iqbal2019actor}.

The critic function $Q_i$ for agent $i$ depends on the embedding $g_i(o_i,a_i)$ of agent $i$ and the attention-weighted average reward of other agents $x_i$ as:
\begin{align}
& Q_i(\bar{o},\bar{a}) = f_i ( g_i(o_i,a_i), x_i)\\
& x_i = \sum_{j \neq i} \alpha_{ij}  \Vcal ( g_j( o_j,a_j) ) \\
&\alpha_{ij} \propto \exp( v_i^T W_i^T W_j v_j)
\end{align}
where $f_i$ and $\Vcal$ are feedforward neural networks.
$\alpha_{ij}$ is the dynamic-attention function given by agent $i$ to agent $j$, and depends on embeddings $v_i = g_i(o_i,a_i)$ and $v_j = g_j(o_j,a_j)$. $W_i$ and $W_j$ are agent specific ``key'' transformation matrices for the attention function. The critic function is learned using a temporal difference loss calculated on trajectories collected using current and past policies, which are stored in a replay buffer $D$. The loss function for updating the critic module for agent $i$ takes following form: 
\begin{multline} 
\Lcal_Q = \sum_{i=1}^{N} E_{(\obar,\abar,r,\obar^{'}) \sim D } \left[ (Q_i(\obar,\abar) - y_i )^2\right] \text{ , where}\\
y_i = r_i + \gamma E_{\abar^{'} \sim \bar{\pi}(\obar^{'}) } \left[ \bar{Q}_i(\obar^{'}, \abar^{'}) - \alpha \log ( \bar{\pi}_i (a_i | o_i) ) \right] 
\label{eq:critic_loss_maac}
\end{multline}
where $\bar{Q}_i$ and $\bar{\pi}_i$ are target critic and target policy for agent $i$. In this work, we build on the multi-agent actor-critic framework described above to incorporate learning of emergent roles in each agent. Next, we describe a popular role learning framework in the cooperative setting.

%% file: 102roma.tex
\subsection{Role-aware policy learning}

Roles have been used in many MARL applications for promoting cooperation and coordination among agents \cite{barrett2015cooperating,roy2020promoting,urieli2011optimizing,zhang2021hierarchical}.
Recently, Wang et al. proposed \roma{,}\cite{wang2020roma}, a MARL framework for learning emergent roles in a cooperative game setting.
In \roma, the roles are designed to exhibit two crucial properties: (1) \textbf{Identifiability: } role should be identifiable with its behavior and diverse enough to fulfill the task, and (2) \textbf{Specialization: } agents with similar roles should specialize in similar responsibilities. For identifiability, it maximizes the mutual information $I( \rho_i ; \tau_i )$ between the role distribution $p(\rho_i|o_i)$, and the trajectory-conditional role distribution 
$p(\rho_i| \tau_i,o_i)$.
For ensuring specialization and diversity, it maintains a dissimilarity model $d_{\phi}$ to measure the distance of trajectories of two agents and seek to maximize $I( \rho_i ; \tau_j ) + d_{\phi} (\tau_i, \tau_j) $. With these regularizers, \roma\ ensures agents adopt role-aware policies within a team in each trajectory.

\roma\ builds on the QMIX~\citep{rashid2018qmix}, which is designed for settings where the agents receive a shared reward for joint observation and actions. \roma\ also works within the Q-learning mode without any additional policy network. It learns per-agent local utility networks, $\Qcal_i$ output of which are passed to a mixing network to compute global TD-loss.
During execution, each agent acts according to optimum policy derived from its' local utility function.
In \roma\, the local utility function $\Qcal_i(o_i,a_i,\rho_i)$ takes the agent's role $\rho_i$ also as input. The role is $\rho_i$ defined as a latent (emergent) variable with a distribution, conditioned on 
the trajectory of the agent $\tau_i$. 
However, the Q-learning framework used in \roma\ cannot be trivially extended to the competitive setting since the composition of local utility functions $\Qcal_i$ into one global utility function $\Qcal$ strongly relies upon the assumption of the common reward of all agents which does not hold in a competitive setting with mixed-rewards. 
Moreover, the local utility function $\Qcal_i$ takes only observation and action of $i$-th agent ($o_i$ and $a_i$) as input, and therefore agents cannot learn from the experience of opponents (other agents with decoupled rewards). 

In the next section, we develop a novel formulation for learning emergent roles in competitive setting in the actor-critic framework, which is better suited towards incorporating mixed-reward.  
In the actor-critic framework, the critic function $Q_i(\bar{o},\bar{a})$ takes observations and actions of all agents as input, hence facilitating the sharing of information among agents with decoupled reward and agents benefits from learning opponent-aware policies by exploiting the shared opponent information available during the training.

%
%


%% file: 103rac.tex
\subsection{Opponent-aware Role learning}
\begin{figure*}[!t]
	\centering
	\subfloat{\includegraphics[width=.8\textwidth]{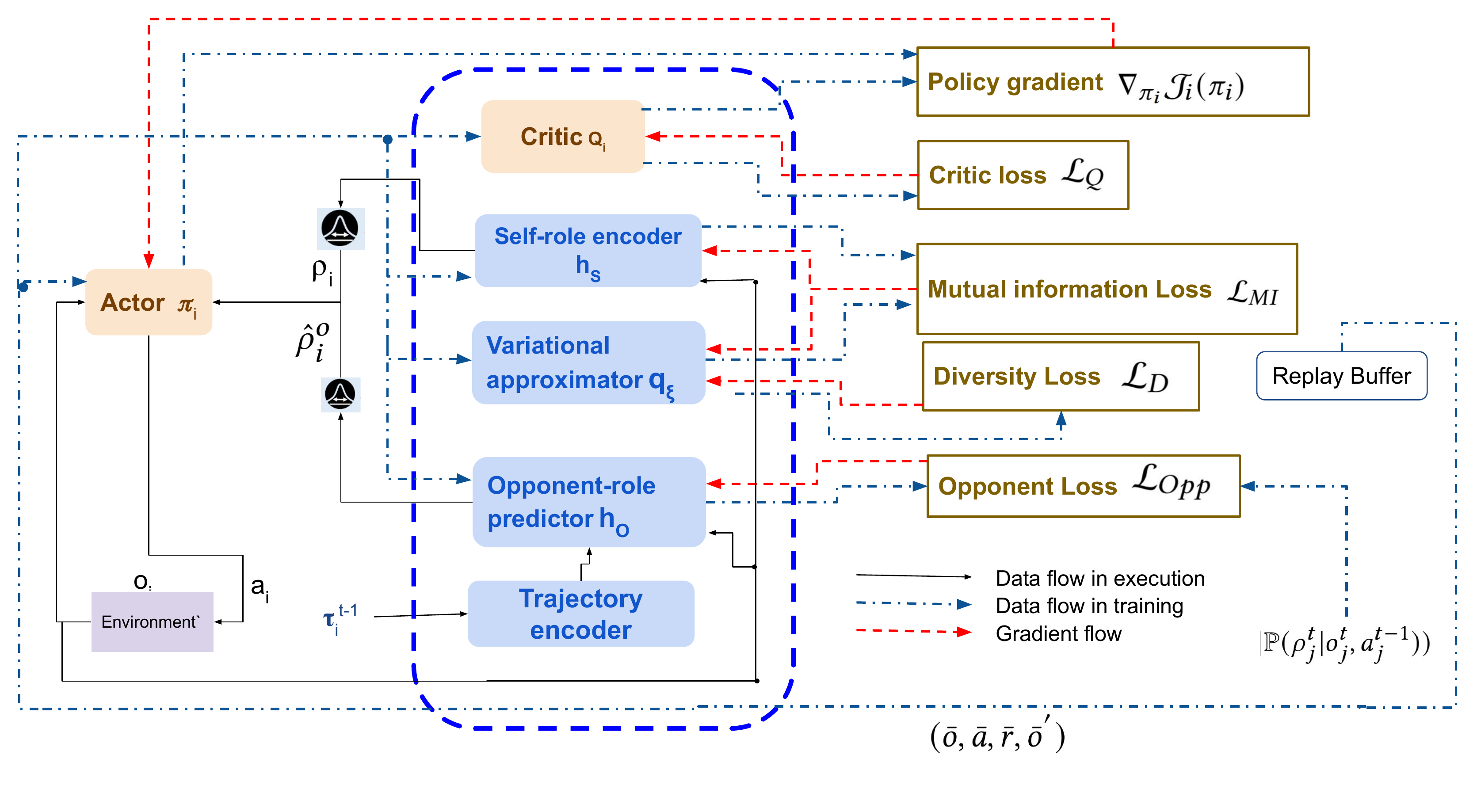}}
	\vspace{-5mm}
	\caption{ A schematic diagram of our approach. 
	The self-role encoder $h_S$ generate the role $\rho_i$ for the $i$-th agent and opponent role predictor $h_O$ predicts roles $\hat{\rho}_{i}^o$ for the opponents. Policy $\pi_i$ takes own observation $o_i$ as well as generated role encodings ($\rho_i, \hat{\rho}_{i}^o$) to generate action $a_i$. 
	Critic $Q_i$ takes both $((o_i, \rho_i, \hat{\rho}_{i}^o), a_i)$ to compute the state-action value. 
	Critic loss trains the critic module, whereas mutual information loss, diversity loss, and opponent loss train the role encoder modules. 
	The framework can be trained in an end-to-end manner. }
	\label{fig:diagram}
\end{figure*}

\begin{algorithm}
	\caption{ \our\ }
	\label{modifiedAlgo}
	Initialize replay buffer, $D$\\
	$T_{update} \leftarrow 0$\\
	\For{$eps = 1 \cdots $ max episodes}{
		$\tau_i^{0} \gets 0$\\
		$D_i \gets \emptyset, \forall i	$\\
		Reset the environment, and get initial $o_i^{1}$ for each agent $i$\\
		\For{$t=1 \cdots $ episode length }{
			Sample roles $\rho^{t}, \hrho^{t}$ using~(\ref{eq:roles}) for all agents\\
			Sample action from respective policies $a_i^{t} \sim \pi_i(o_i^{t},\rho_i^{t},\hrho^{o,t}_i)$ for each agent $i$\\
			Each agent receives $o_i^{t+1}, r_i^{t}$ from environment\\
			$\tau_i^{t} \gets \tau_i^{t-1} + (o_i^{t},a_i^{t})$\\
			$D_i \gets D_i + (o_i^{t},\tau_i^{t-1},a_i^{t},r_i^{t},o_i^{t+1})$\\
		}
		$T_{update}=T_{update}+1$\\
		\If{$T_{update} \geq $ min steps per update}{
			\For{$j=1 \cdots $number of updates}{
				$H_B \gets B$ episodes sampled from $D$\\
				Update($H_B$)\\
			}
			
			$T_{update} \leftarrow 0$\\
		} 		 
	}
\end{algorithm}

\begin{algorithm}
	\caption{Update($H_B$)}
	\label{update}
	Select $t_1,\cdots, t_B$, the starting time-step for each of the $B$ episodes\\
	$\Lcal_{MI} \gets 0,\Lcal_{D} \gets 0, \Lcal_{Opp} \gets 0$\\
	\For{$j = 1 \cdots \text{episode length}$ }{
		$S \gets \emptyset$ \\
		\For{$k=1 \cdots B $}{
			$S_k \gets (\obar^{t_k+j,k}, \bar{\tau}^{t_k-1+j,k}, \abar^{t_k+j,k}, \rbar^{t_k+j,k}, \obar^{t_k+1+j,k})$\\
			$S \gets S  + S_k$\\
		    $\rho^{j,k},\hat{\rho}^{j,k},\Lcal_{MI}^{j,k},\Lcal_{D}^{j,k}, \Lcal_{Opp}^{j,k} \gets Role(S_k)$\\
    		$\Lcal_{MI} \gets \Lcal_{MI} + \Lcal_{MI}^{j,k}$\\
    		$\Lcal_{D} \gets \Lcal_{D} + \Lcal_{D}^{j,k}$\\
    		$\Lcal_{Opp} \gets \Lcal_{Opp} + \Lcal_{Opp}^{j,k}$\\
		}
    Update critics, policies and targets using $S, \rho^{j,\cdot}, \hat{\rho}^{j,\cdot}$\\
	}
	$\psi_{Role} \gets \argmin_{\psi}(\Lcal_{MI}+\Lcal_{D}+\Lcal_{Opp})$
\end{algorithm}
\begin{algorithm}
	\caption{Role($S$)}
	\label{role}
	Obtain roles $\rho^{t}, \hrho^{t}$ from eq. (\ref{eq:roles}) \\
	Calculate $\Lcal_{MI}$, $\Lcal_{D}$ and $\Lcal_{Opp}$ using (\ref{eq:miloss}), (\ref{eq:dloss}), and (\ref{eq:opp_loss})\\
	return $\rho^t,\hat{\rho}^{t},\Lcal_{MI},\Lcal_{D},\Lcal_{Opp}$
\end{algorithm}

We introduce \our, the proposed a role-based actor-critic algorithm,  designed for opponent-aware learning in the team-competitive games. In this setting, the agents in the team receive individual as well as shared rewards while the agents in different teams do not receive any shared reward. In this section, we describe the algorithm for simplicity the setting with two teams, each with any number of cooperating agents, which compete with each other according to the rules of individual games (described in section \ref{sec:experiments}). An interesting question in  this setting is to determine whether learning opponent-aware role-based policies is beneficial for the teams. 
	
\our\ is built on the \bsln\ framework described above, and uses both agent-specific policy network $\pi_i$ and critic network $Q_i$. Additionally \our\ employs two other networks:
\begin{enumerate}
	\item  $h_S$: self role encoding network
	\item $h_O$: opponent role prediction network
\end{enumerate}

Another recurrent neural network (GRU) is used to maintain a running encoding of the trajectory $\tau_i^t$ at time $t$. For simplicity, we overload the notation $\tau_i^t$ to denote the embedding of the trajectory till time $t$, for the rest of the paper. Intuitively, the policy of an agent $i$ depends on its observation $o_i$, latent role $\rho_i$, and predicted roles of opponent agents $\hrho_i^{o} = [\hrho_j]_{j \in Opp(i)}$, where $Opp(i)$ is the set of opponent agents of agent $i$. Hence, $\pi_i(o_i,\rho_i,\hrho_i^o)$ computes the probability of action $a_i$. The self role encoding network $h_S$ and opponent role prediction network $h_O$ take the current observation $o_i^t$ and past action $a_i^{t-1}$ for $h_S$, or trajectory encoding $\tau_i^{t-1}$ for $h_O$ as input; and output a Gaussian distribution over the respective roles:
\begin{align}
&( \mu_{\rho_i^t}, \sigma_{\rho_i^t})= h_S( o_i^t, a_i^{t-1} )\ ;\ \ \PP(\rho_i^t | o_i^t, a_i^{t-1}) = \Ncal ( \mu_{\rho_i^t}, \sigma_{\rho_i^t}) \\
&\hat{\mu}_{\rho_j^t},\hat{\sigma}_{\rho_j^t} = h_O( o_{i}^t, \tau_{i}^{t-1} ) \forall j\neq i\ ;\ \ \hat{\PP}( \hat{\rho}_j^t |  o_{i}^t, \tau_{i}^{t-1}) = \Ncal ( \hat{\mu}_{\rho_i^t}, \hat{\sigma}_{\rho_i^t}) 
\label{eq:roles}
\end{align}
Finally, the critic network $Q_i$ for agent $i$ takes as input the joint observation $\bar{o}$, joint action $\bar{a}$, and the role embeddings of current agent $\rho_i$ and role predictions of its opponents $\hrho_i^o$, $Q_i(\bar{o},\rho_i,\hrho_i^o,\bar{a})$, and outputs an estimate of the value-to-go. Figure \ref{fig:diagram} shows the overall architecture of the \our\ system. 

In the network architecture described above, the per-agent policies $\pi_i$ are trained using standard policy gradient steps, whereas the parameters of critic network $Q_i$, self-role encoder network $h_S$, and opponent-role predictor network $h_O$ are updated by centralized training of critics. 
To update policy network of agent $i$, the policy gradient in~(\ref{eq:pol_grad_maac}) is further modified as follows: 
\begin{multline}
  \nabla_{\pi_i} \Jcal_i(\pi_{i}) = E_{\obar \sim D,\abar \sim \pi,\rho \sim h_S, \hrho \sim h_O} \left[ \nabla_{\pi_i} \log (\pi_{i} (a_{i}|o_{i},\rho_i,\hrho_i^o) \right.\\
   \left. ( Q_i(\bar{o},\rho_i, \hrho_i^o, \bar{a}) 
   - b(\obar,\rho_i, \hrho_i^o, \abar_{-i}) 
   - \alpha \log (\pi_i (a_i | \obar_i, \rho_i, \hrho_i^o))) \right]
\end{multline}
Correspondingly, baseline computation is also modified as: 
\begin{multline}  
    b(\obar,\rho_i, \hrho_i^o, \abar_{-i}) 
    = \EE_{a_i \sim \pi_i(o_i, \rho_i, \hrho_i^o)}  \left[ Q_i(\obar,\rho_i, \hrho_i^o,(a_i,\abar_{-i}))\right] \\
    = \sum_{a_i \in A_i} \pi_i(a_i |  o_i, \rho_i, \hrho_i^o ) Q_i(\obar,\rho_i, \hrho_i^o,(a_i,\abar_{-i}))
\end{multline}
While parameters of $\pi_i$ are trained using policy-gradient updates, parameters of the other networks are learned jointly by minimizing an aggregate loss function described next.
Firstly, the critic loss $\Lcal_Q$ defined in~(\ref{eq:critic_loss_maac}) is modified as follows:
\begin{multline} 
	\Lcal_{Q}=\sum_{i=1}^N  \EE_{(\obar,\abar,\rbar,\obar^{\prime}) \sim D,\rho_i \sim h_S, \hrho_i^{o} \sim h_O } \left[ (Q_i(\obar ,\rho_i,\hrho_i^o, \abar ) - y_i )^2 \right] \text{ , where} \\
    y_i = r_i + \gamma \EE_{\abar^{\prime} \sim \bar{ \pi }, \rho_i^{'} \sim h_S, \hrho_i^{o,'} \sim h_O } 
    \left[ \bar{Q}_i(\obar^{'}, \rho_i^{'}, \hrho_i^{o,'},\abar^{'}) \right. \\
    \left. - \alpha \log ( \bar{\pi}_i (a_i^{'} | o_i^{'}, \rho_i^{'}, \hrho_i^{o,'}) ) \right] 
\label{eq:critic_loss_role}
\end{multline}

\begin{figure*}[!t]
	\centering
	\subfloat[\ourgame]{\includegraphics[ width=.35\textwidth]{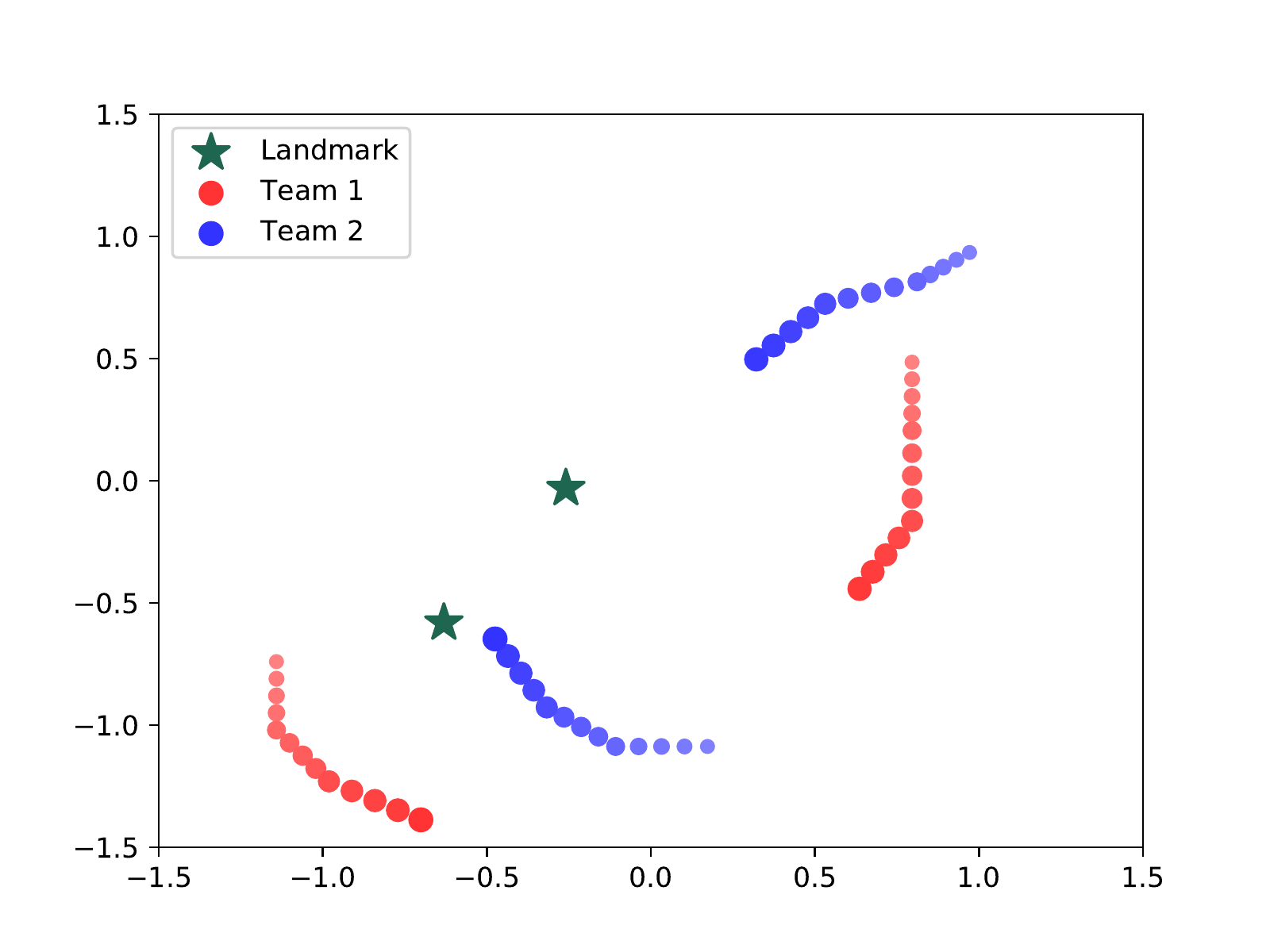}}\hspace{3mm}
	\subfloat[\market]{\includegraphics[width=.35\textwidth]{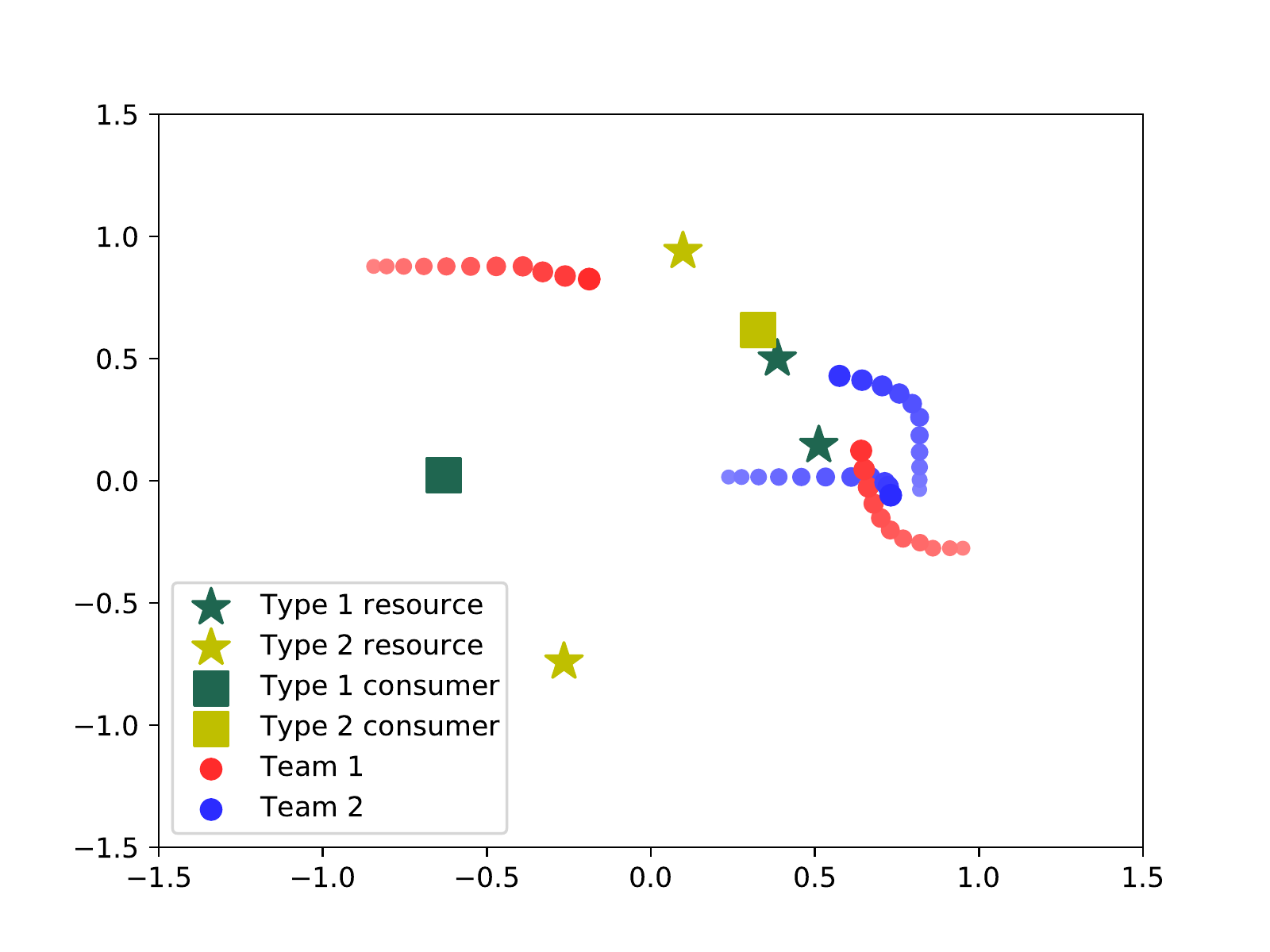}}
	\caption{\ourgame~(left) is a two-dimensional board game where members of each team try to reach one of the landmarks before any agent from the opponent team reaches any landmark. In \market~(right) teams pick resources scattered on the board and drop those to respective consumers before the opponent does the same.}
	\label{fig:evironments}
\end{figure*}

Now, we describe the losses employed for training the self and opponent role encoder modules. For maintaining identifiability of roles with trajectories, we like to minimize conditional entropy $H(\rho_i^t \vert o_i^t, \tau_i^{t-1} )$, which is often intractable. Therefore, we maintain a variational approximation of distribution over roles, conditioned upon trajectories: $q_{\xi}(\rho_i^t | o_i^t, a_i^{t-1}, \tau_i^{t-1})$ and maximize the mutual information between $q_{\xi}(\rho_i^t |  o_i^t, a_i^{t-1}, \tau_i^{t-1})$ and $\PP(\rho_i^t |o_i^t, a_i^{t-1})$. Simultaneously we minimize the entropy of role distribution $\PP(\rho_i^t |o_i^t, a_i^{t-1})$ to encode the history in the role encoder network.
Hence, corresponding mutual information loss, $\Lcal_{MI}$, is defined as:
\begin{multline} \label{eq:miloss}
 \mathcal{L}_{MI} = 
 \sum_{i=1}^N \EE_{\tau_i^{t-1},o_i^t \sim D} \left[ D_{KL}(\PP(\rho_i^t|o_i^t,a_i^{t-1})|| \right. \\
\left. q_{\xi}(\rho_i^t|o_i^t,a_i^{t-1},\tau_i^{t-1})) \right] 
 + H ( \PP(\rho_i^t|o_i^t,a_i^{t-1}) ) 
\end{multline}
Here, $D_{KL}$ denotes KL-divergence,
and $H(\cdot)$ denotes the entropy function. 
In order to promote diversity between the behaviors of agents within a team, we introduce a new diversity loss $\mathcal{L}_D$, defined as: 
\begin{equation}\label{eq:dloss}
	\Lcal_D = \sum_{k \in \Kcal } \sum_{(i,j) \in \Ncal(k)} 
	\EE_{(\bar{\tau}^{t-1},\obar^t) \sim D}
	\left[ q_{\xi}(\rho_i^t|\tau_j^{t-1},o_j^t,a_i^{t-1}) \right]
\end{equation} 
Here, $\Kcal$ is the set of teams and $\Ncal(k)$ are agents in team $k$. The loss penalizes the probability of $\rho_i$ being identified with trajectory $\tau_j$ of another agent in the same team.
%
Finally, to train the opponent role prediction network $h_O$, we define the opponent loss $\Lcal_{Opp}$ as:
\begin{equation}\label{eq:opp_loss}
\Lcal_{Opp} = \sum_{\substack{i,j:team(i) \\ \neq team(j)}}
\EE_{(\bar{\tau}^{t-1},\obar^t) \sim D}
\left[ D_{KL}(\hat{\PP}(\rho_j^t|o_{i}^t,\tau_i^{t}) || \PP(\rho_j^t|o_j^t,a_j^{t-1})) \right]
\end{equation} 
This loss penalizes the divergence of $\hat{\PP}(\rho_j^t|o_{i}^t,\tau_{i}^t)$, the predicted opponent role distribution for agent $j$ from the point of view of agent $i$,  from  $\PP(\rho_j^t|o_j^t,a_j^{t-1})$, the self-role distribution of agent $j$.
We define the composite loss function for training the role networks as:
\begin{equation}
\Lcal_{Role}=\Lcal_{D}+\Lcal_{MI}+\Lcal_{Opp}
\end{equation} 
Hence, $\Lcal_Q + \Lcal_{Role}$ is the total loss used to learn the combined critic module of \our\ model. However, it is observed that the evolution of the role networks in the later phase of training causes instability in learning policies, which result in reduced average reward. This phenomenon is mitigated by using an exponentially decaying weightage on role loss $\Lcal_{Role}$. The total critic loss becomes:
\begin{equation}
\Lcal_{tot}=\Lcal_Q  + \lambda^{\frac{u}{C}} \times \Lcal_{Role}
\end{equation}
 where $\lambda\in [0,1]$ is the decay factor, $ u $ is the training episode count, and $ C $ is a constant affecting the rate of decay over episodes. Algorithm~\ref{modifiedAlgo} includes a pseudo-code for our algorithm. Table ~\ref{tab:notations} contains list of all variables. 

%% file: 200exp.tex
\section{Experimental Results}
\label{sec:experiments}

\begin{figure*}[!t]
	\centering	
	\subfloat{\includegraphics[width=.23\textwidth]{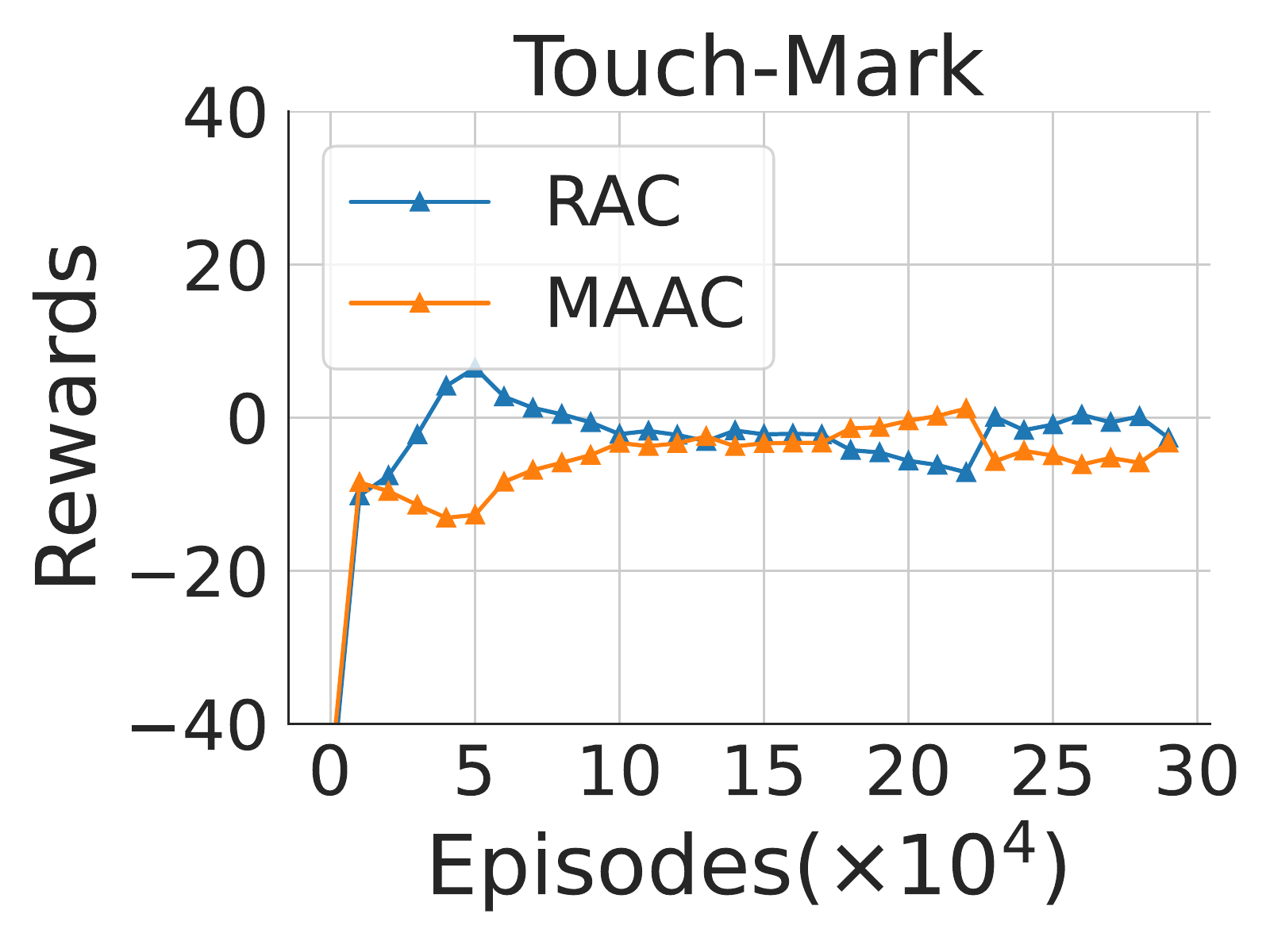}}
	\subfloat{\includegraphics[width=.23\textwidth]{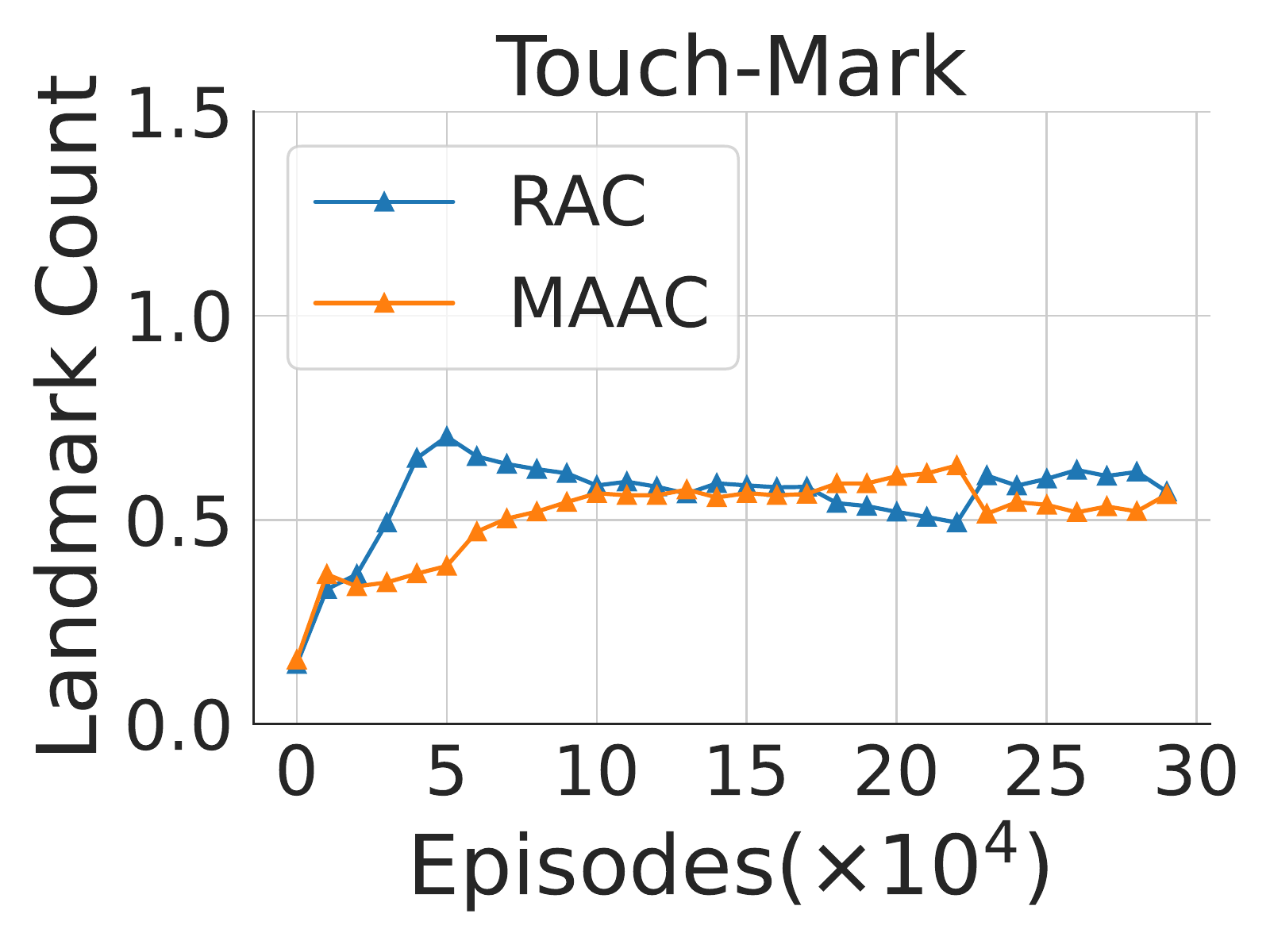}}
	\subfloat{\includegraphics[width=.23\textwidth]{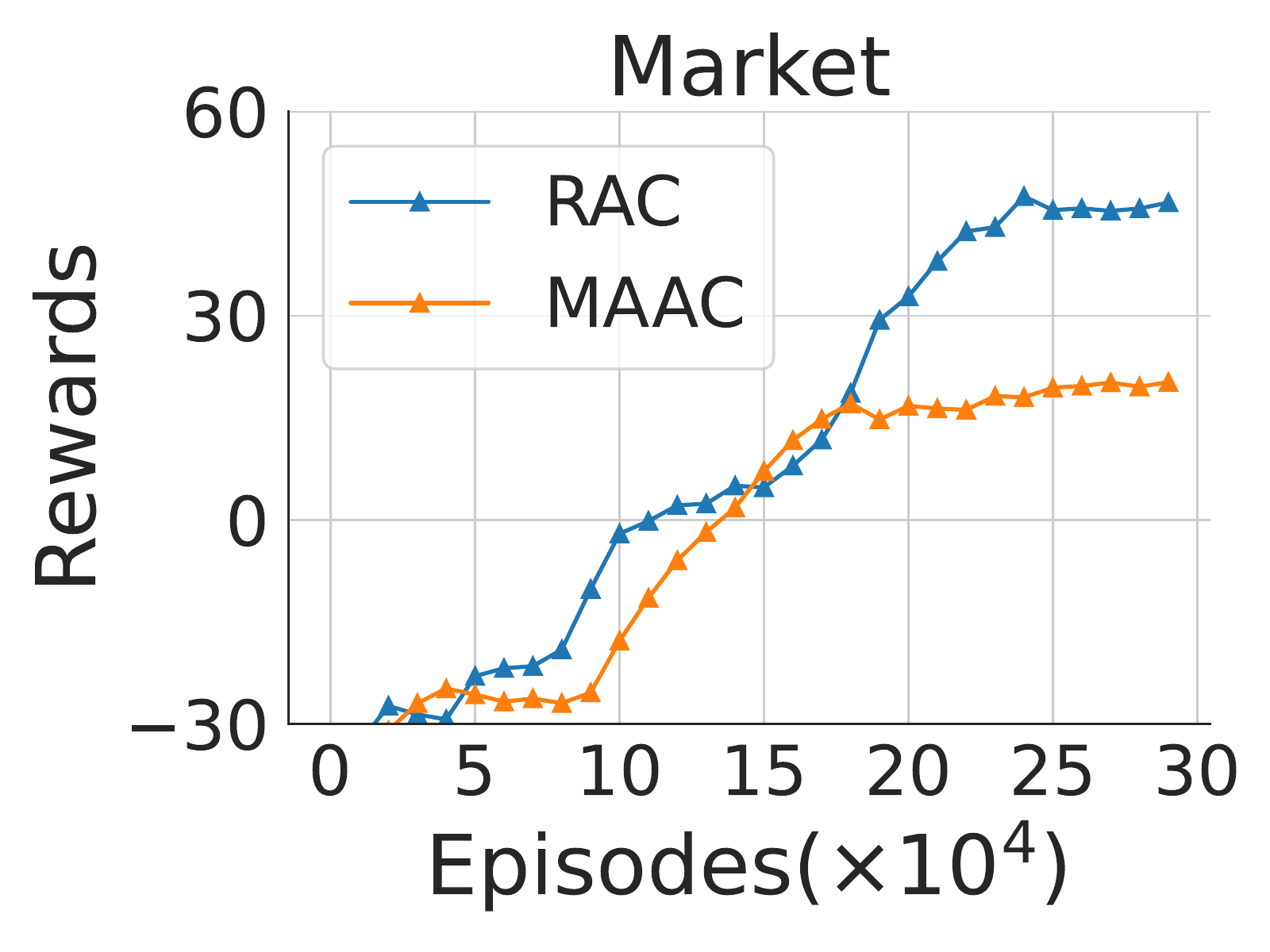}}
	\subfloat{\includegraphics[width=.23\textwidth]{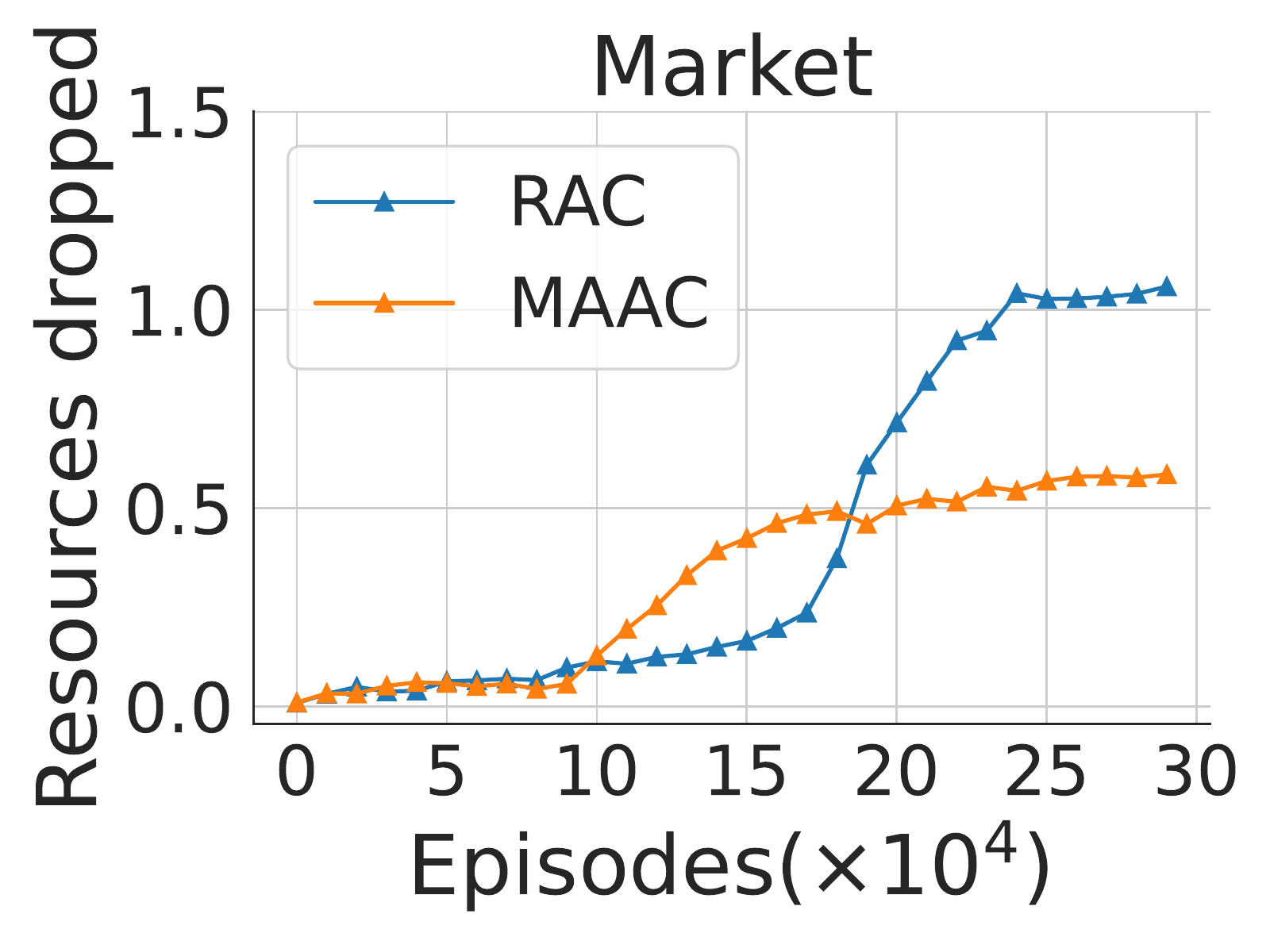}}	
	\caption{Performance comparison. We play \our\ agents against \bsln\ agents, who are separately trained using self-play for $300000$ training steps. We play them against each other after each $1000$ step and report performance, averaging over $4$ pairs of tournament-play. For \ourgame, we report the average team reward and the fraction of times each team reaches the landmark. For \market, we report the average team reward and the fraction of times each team drops a resource to its respective consumer, after picking it from its source. }
	\label{fig:tournament}
\end{figure*}

\begin{figure}[!t]
    \centering	
	\subfloat{\includegraphics[width=0.23\textwidth]{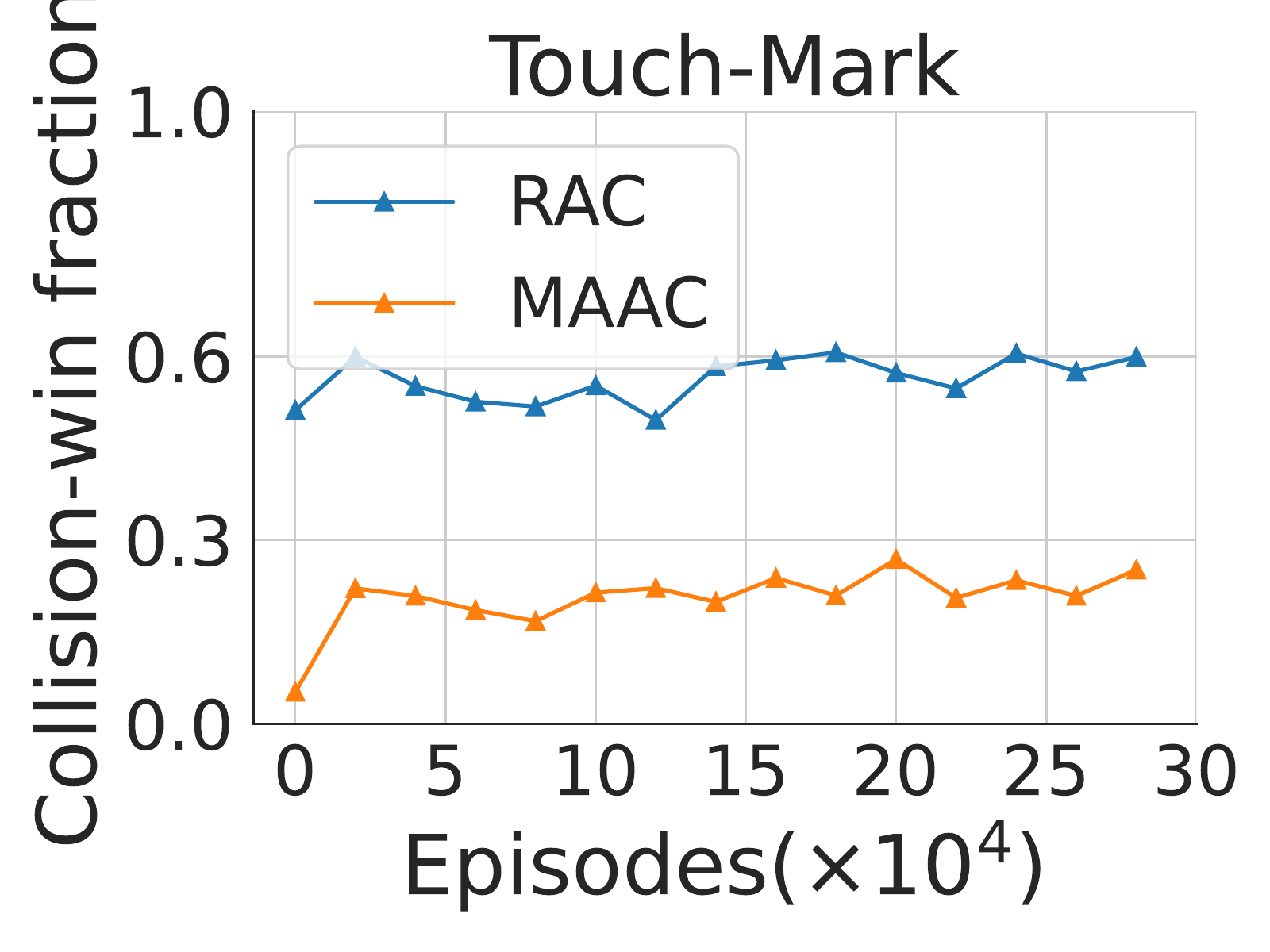}\label{fig:role-catch-goal}}
	\subfloat{\includegraphics[width=0.23\textwidth]{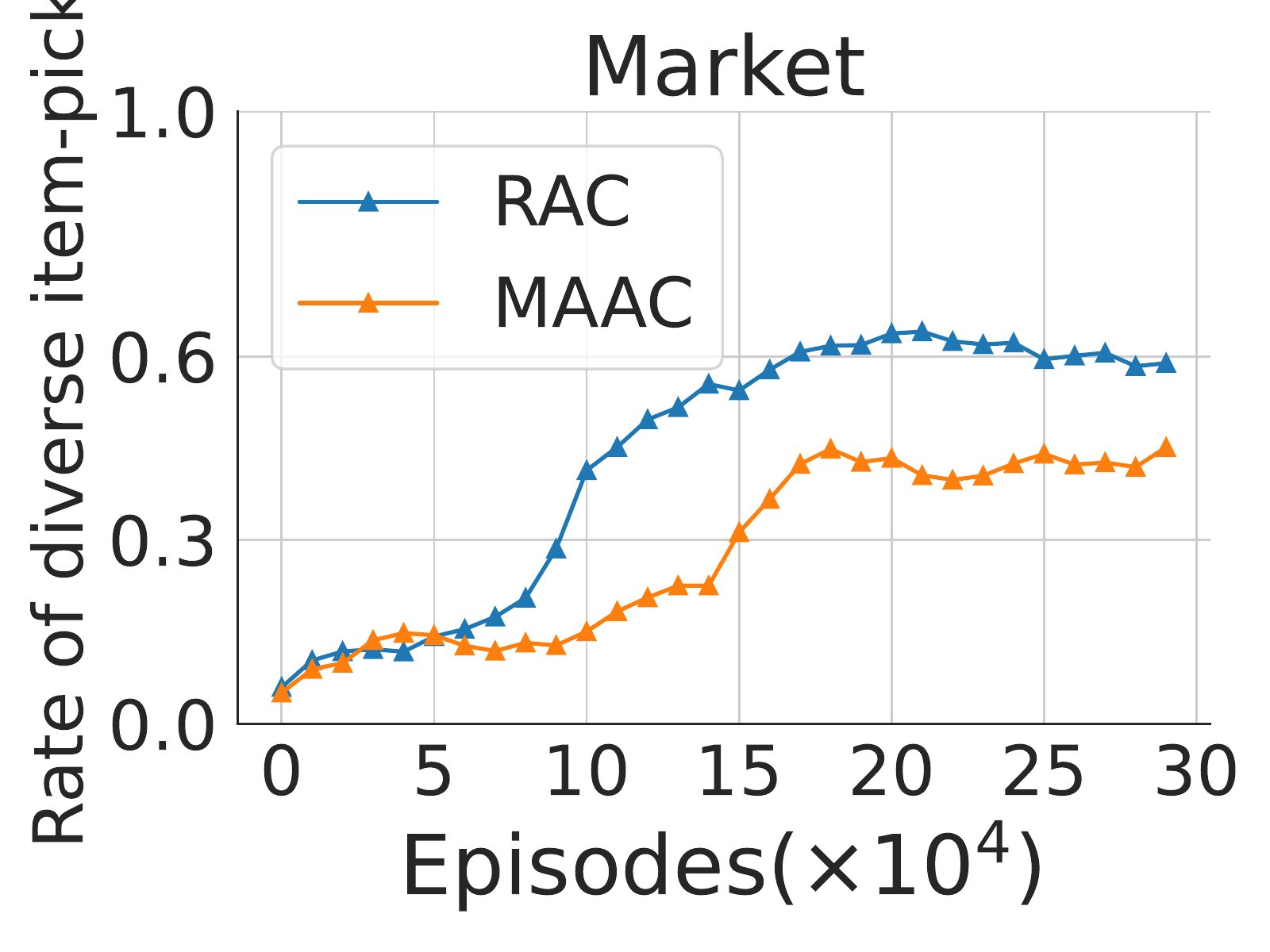}\label{fig:role-market}}
	\caption{Role emergence. For \ourgame, we observe, \our\ agents result in more collisions than \bsln\ agents. In \market, we observe that \our\ agents choose to pick more diverse items than \bsln\ agents. }
	\label{fig:role}
\end{figure}

\begin{figure}[!t]
	\centering
	\subfloat{\includegraphics[width=0.23\textwidth]{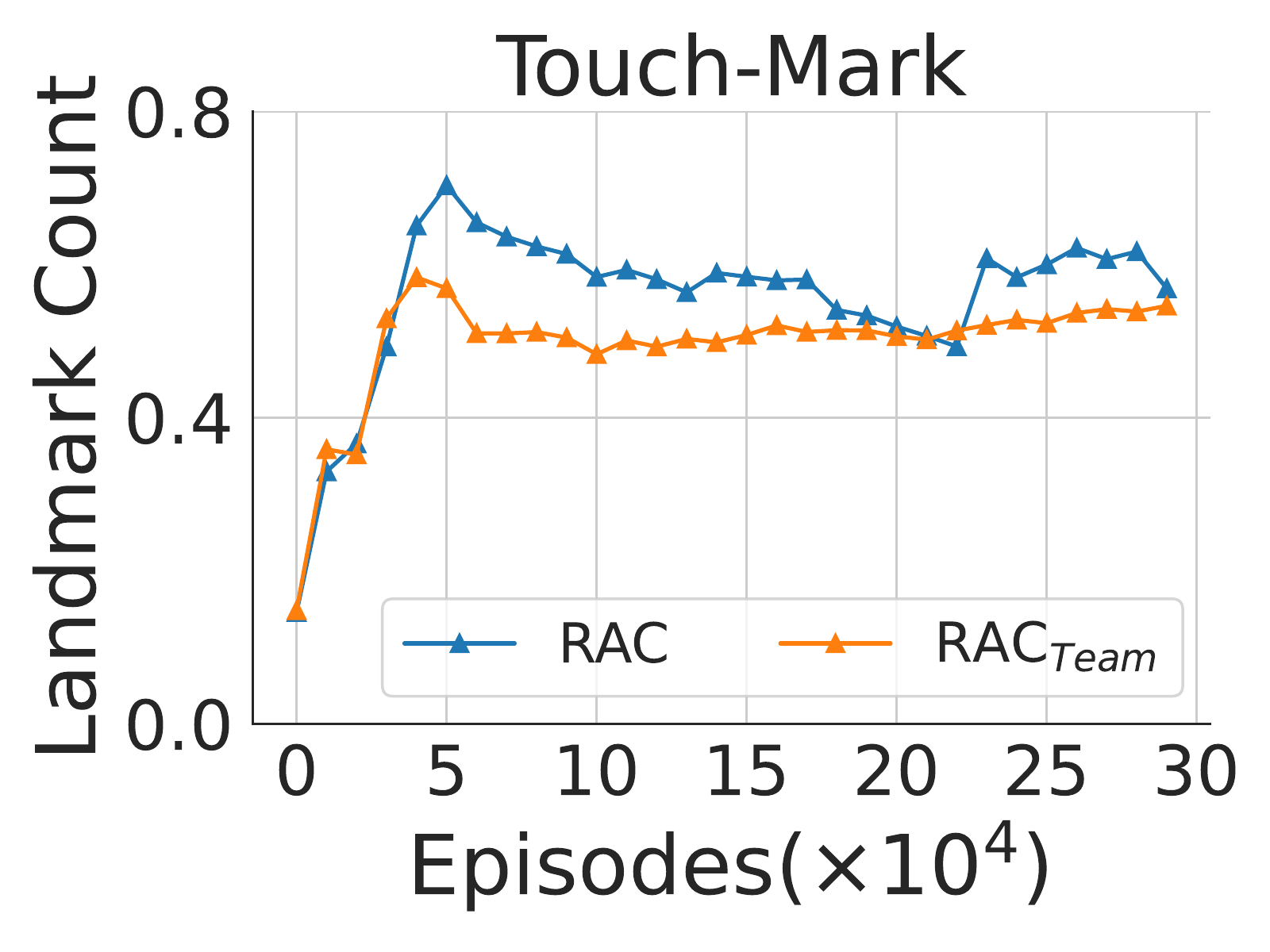}\label{fig:cooperative-catch-goal}}
	\subfloat{\includegraphics[width=0.23\textwidth]{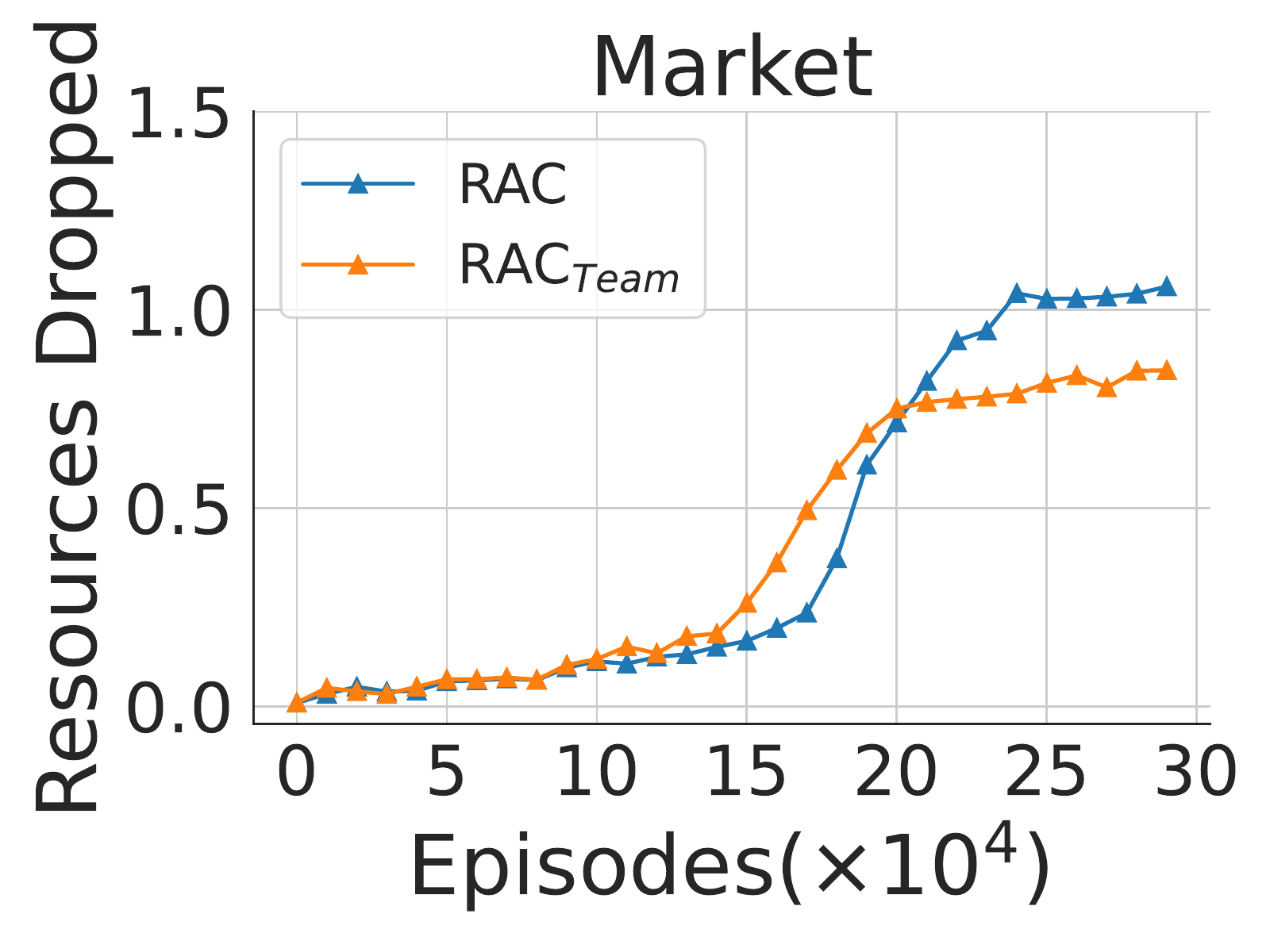}\label{fig:cooperative-market-2}}
	\caption{We compare \our\ with \bslncoop, an extension of \roma\ with opponent modeling removed. However, we observe that performance-wise \our\ is still outperforming \bslncoop . }
	\label{fig:cooperative}
\end{figure}

\begin{figure*}[!t]
	\centering
		\subfloat[]{\includegraphics[width=.23\textwidth]{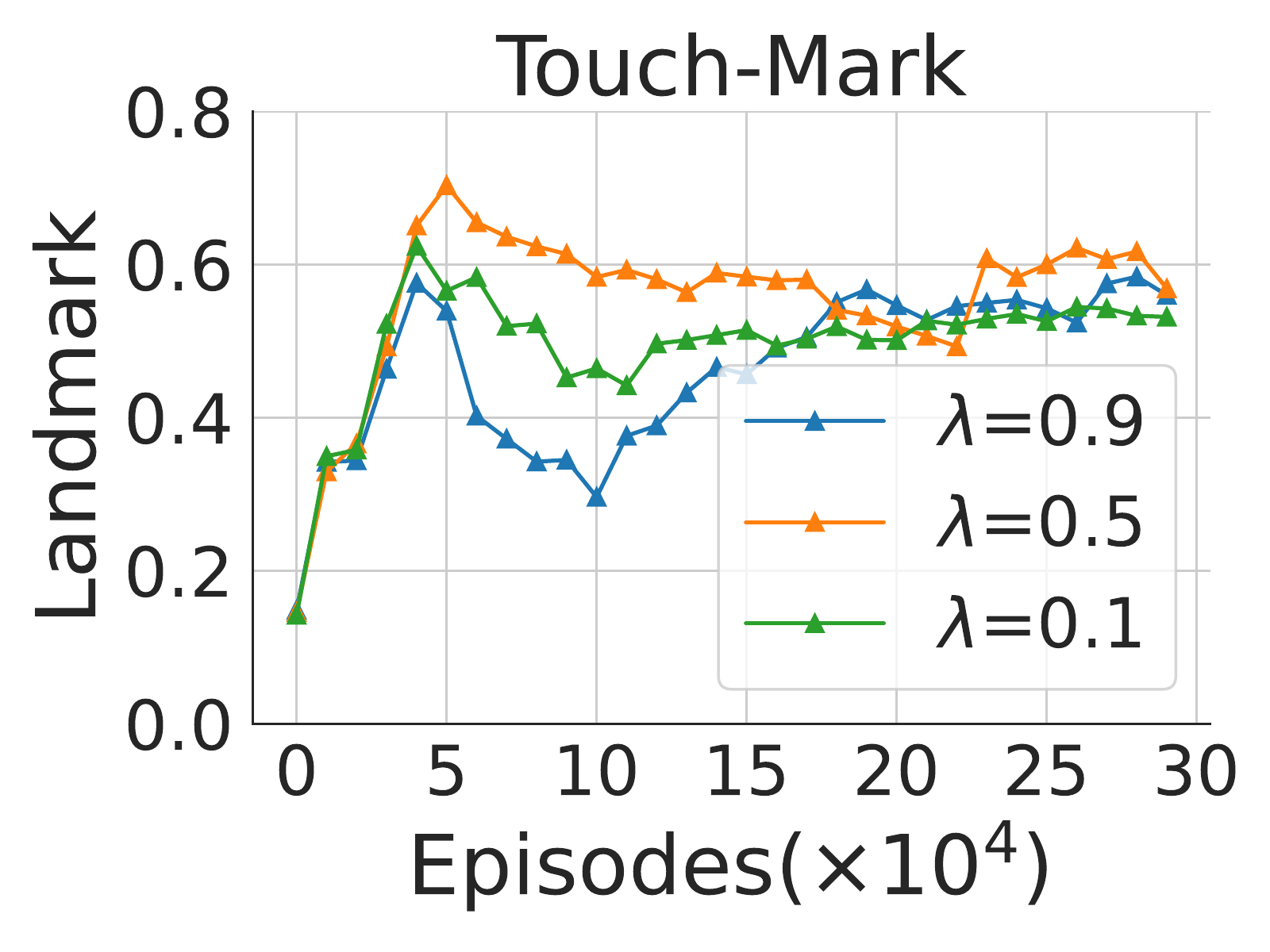}\label{fig:vary_decay_catch_goal}}
		\subfloat[]{\includegraphics[width=.23\textwidth]{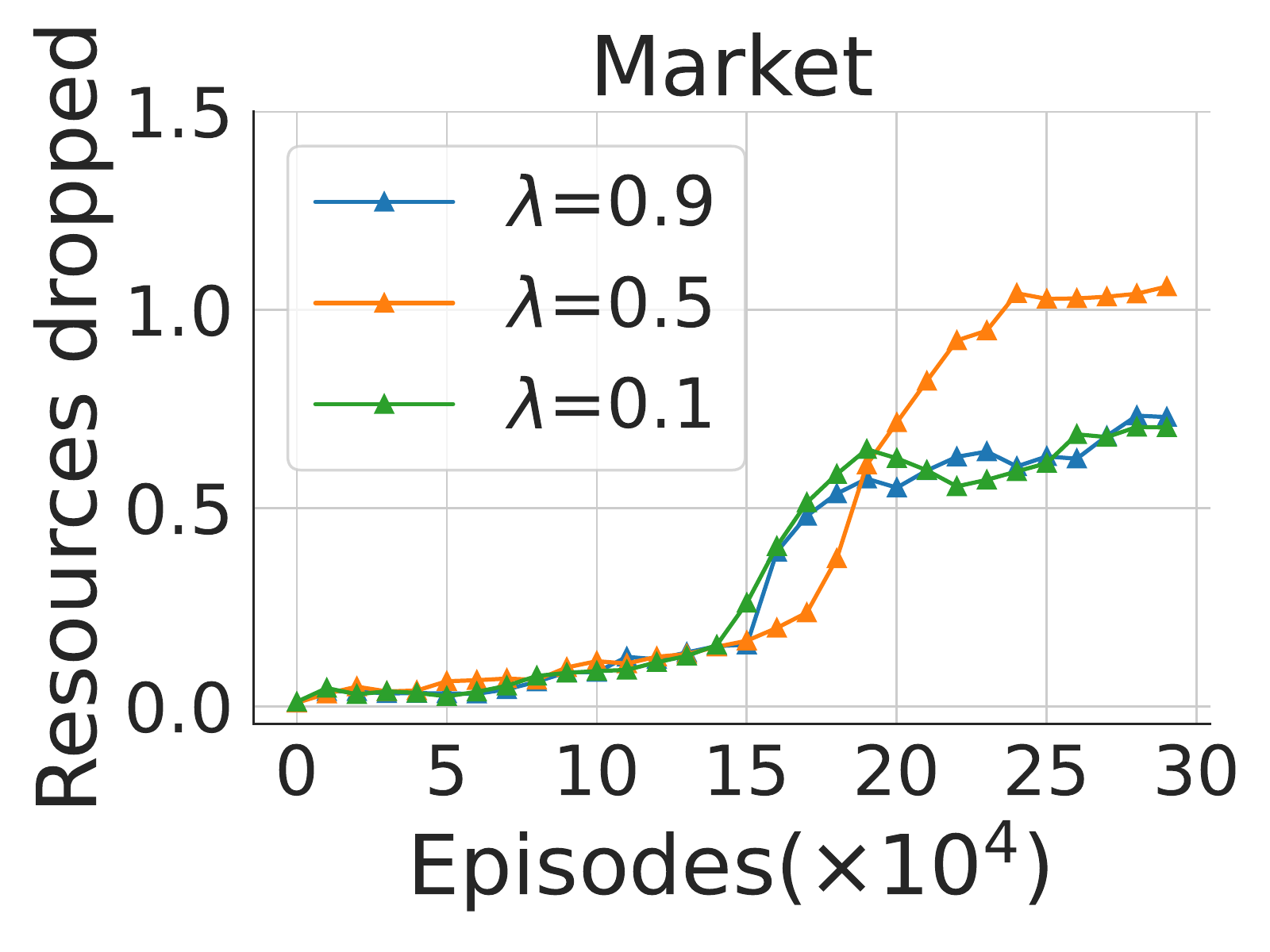}\label{fig:vary_decay_market}}
		\subfloat[]{\includegraphics[width=0.23\textwidth]{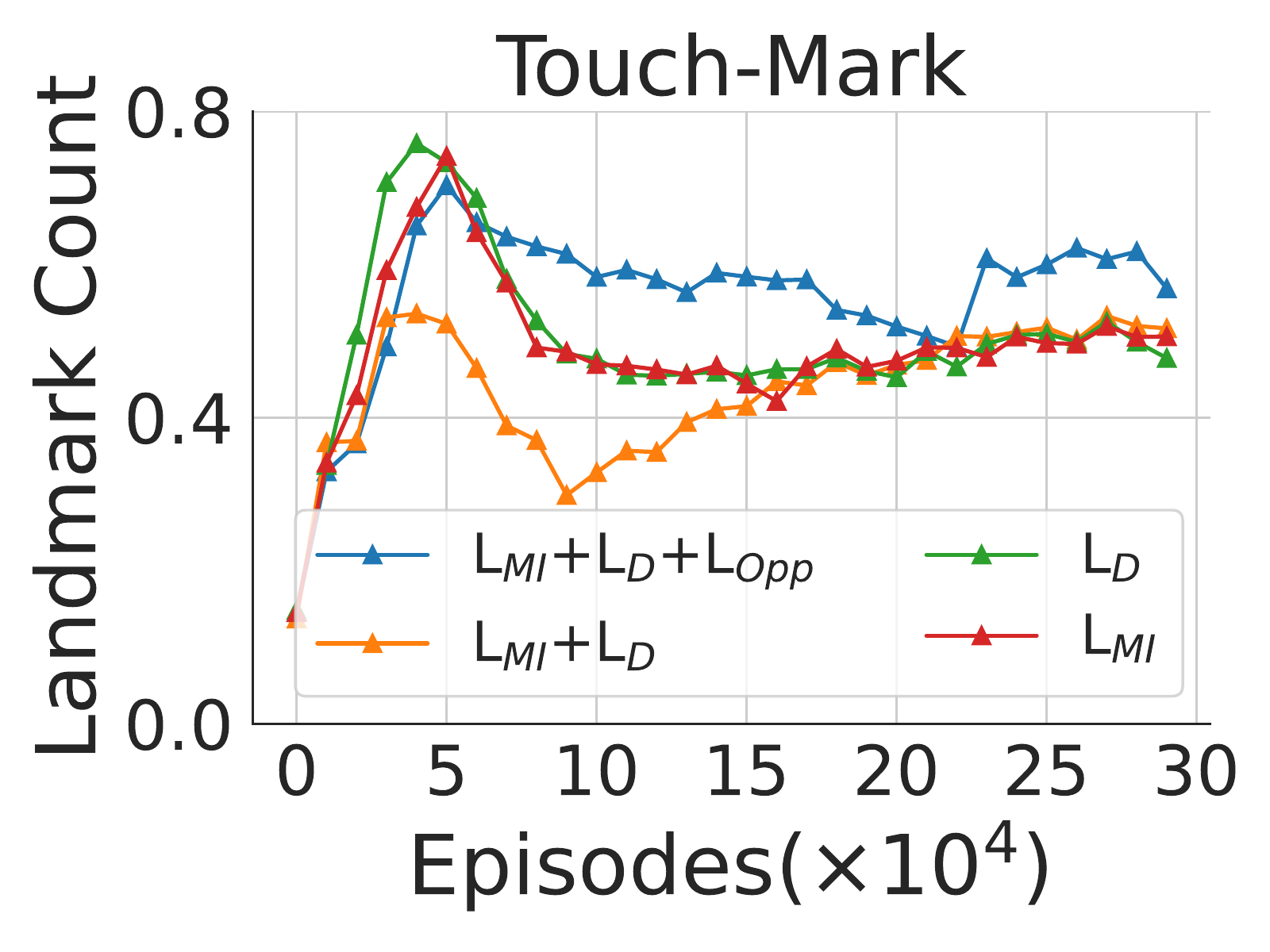}\label{fig:ablations_catch_goal}}
		\subfloat[]{\includegraphics[width=0.23\textwidth]{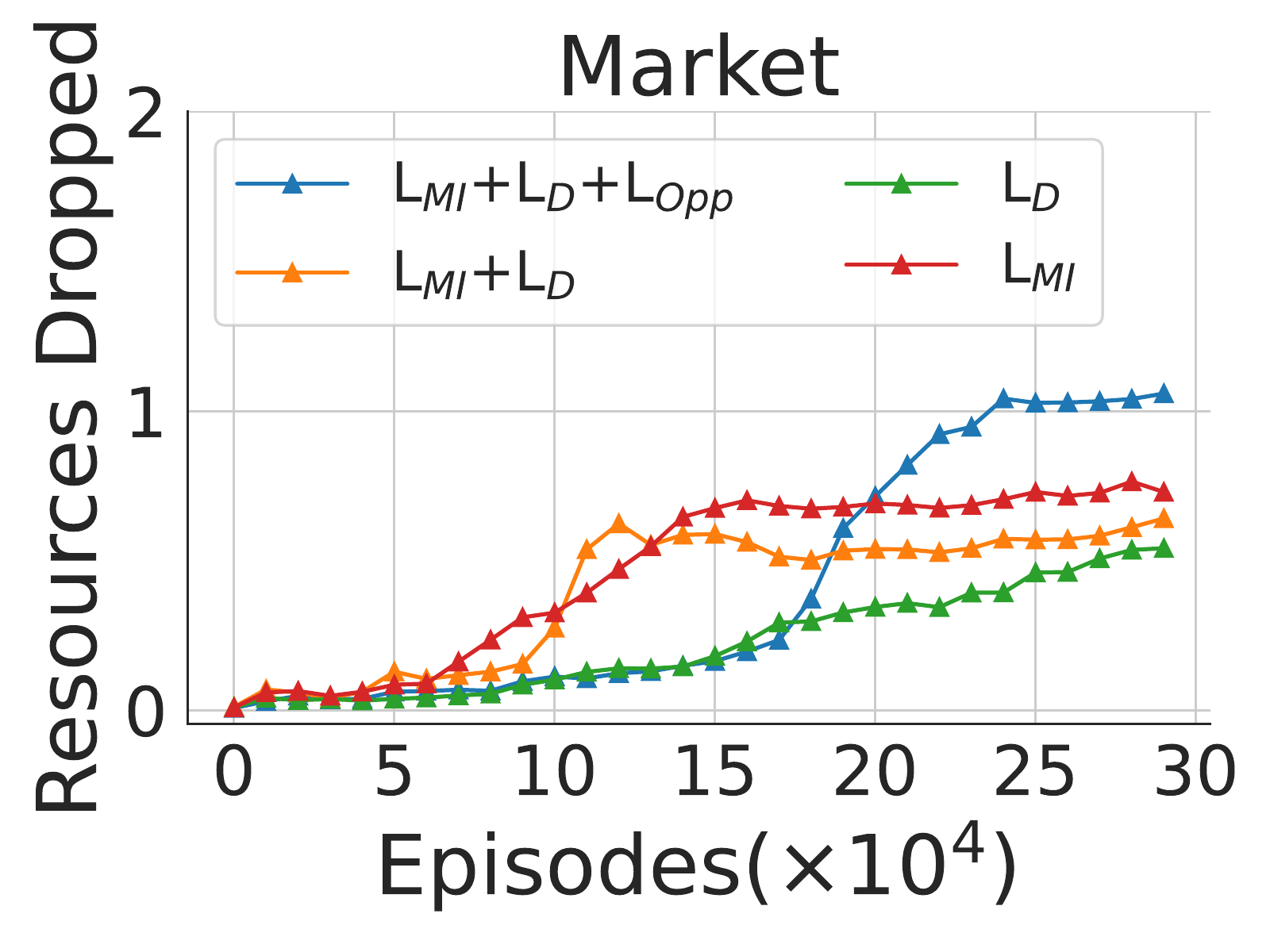}\label{fig:ablations_market}}
		\caption{(Left) Hyperparameter tuning. Here the decay parameter $\lambda$ is varied in $[0,1]$ and the performance of \our\ teams, played against a random set of \bsln\  models, is reported. (Right) Ablations. Performance of various ablations of \our\ are reported, which show that the three distinct objectives are necessary for best performance in both the environments.}
\end{figure*}

In this section, we investigate two primary objectives. First, we check how employing the role network influences the performance of our algorithm in direct competition with baseline algorithm. Second, we check the effectiveness of the role  network in emerging roles dynamically within the team. 
Towards this end, we employ two games, namely \ourgame\ and \market, designed in MPE~\citep{lowe2017multi,mordatch2018emergence}.

\noindent \textbf{\ourgame~\citep{koley2022offsetting}{:}}
It is an episodic board game that consists of two teams, each comprising of multiple agents (2 in our case) and two static landmarks, all randomly positioned at the beginning of each episode and each team tries to reach at least one of those landmarks earlier than  any member of the other team.
The agents start with velocity zero and slowly accelerate until they reach maximum permissible velocity.
The episode ends when an agent reaches a landmark. The winning agent's team  (i.e. both the team members) receives a large reward $r_l$, incurring $-r_l$ penalty to the opposite team. 
Additionally, each agent receives a small penalty,  
proportional to its distance from 
the nearest target at every time step to guide it towards the nearest target.
Moreover, an agent can collide with an agent of the opponent team to divert it from its path. The collision makes both agents temporarily stationary. This mechanism is introduced so that an agent 
has the option to stop an agent of the opponent team from reaching a landmark, thus facilitating the  fellow teammate to  reach a landmark first.

\noindent \textbf{\market{:}}
This is a novel episodic board game with two teams, each consisting of multiple players ($2$ in our case).
At the beginning of each episode, four resources, two for each type of resource are randomly placed on the board, along with one consumer for each type of resource. 
Agents need to acquire the resources and drop those to the respective consumers.
%
After acquiring a resource, the agent continues to receive a penalty proportional to its distance from the appropriate consumer, to guide him towards the corresponding consumer. Both acquiring and dropping resources result in high rewards for corresponding team. 
The termination condition for the episode is timeout of 50 steps.
The consumer remains alive only until one resource of its type is dropped to it. Therefore for highest scores, one team must acquire both types of resources and drop them to respective consumers before their opponents. 

\noindent \textbf{Experimental setup}
We use a one-layer feed-forward neural network followed by GRU unit for getting initial embedding of the trajectories. For each team, a double-layer MLP is employed for each of $\PP(\rho_i^t | o_i^t, a_i^{t-1})$,$\hat{\PP}(\rho_j^t | o_i^t, a_i^{t-1})$ and $q_{\xi}(\rho_i^t | \tau_i^{t-1},o_i^t)$. Critic and policy network architecture follows MAAC~\citep{iqbal2019actor} except inputs of both critic and policy modules for each agent $i$ are modified to include a role tuple for that agent, containing self-role $\rho_i^t$ and predicted opponent-roles $\hrho_i^{o,t}$. 

\noindent \textbf{Baselines}
Our primary baseline is MAAC~\citep{iqbal2019actor}. 
We also compare our method with \bslncoop\ which follows \our\ except does not share information across teams during the training phase. 
This mimics the setting where competing teams separately employ a cooperative role-oriented algorithm like \roma\ and only sacrifices the advantage of training in presence of opponents. 

\noindent \textbf{Metrics}
We report cumulative team rewards per episode to measure the performance of the algorithms. As a qualitative measure, we report the rate of touching landmarks per episode for each team in \ourgame, and the average number of dropped resources per episode for each team in \market{.} The average resources dropped per episode can vary between $0$ and $2$ as there are $2$ resources in \market. All measures are reported after taking an average across all seeds used. 

%

\subsection{Results}
\noindent \textbf{Direct tournament:} Here, we compare \our\ with \bsln\ by directly playing them against each other. Here, we use the \our\ and \bsln\ teams, trained in self-play for $3 \times 10^5$ training episodes starting with random seed. At fixed interval during the training we play them among themselves for $1000$ episodes. This experiment is repeated for $4$ pairs of \our\ and \bsln\ teams, and the average metrics are reported in fig. \ref{fig:tournament} for \ourgame~(left) and \market~(right){.} 
We observe that, as the training progresses, \our\ continues to outperform \bsln. The performance gap is quite significant for \market, whereas, in \ourgame\ the gain is marginal. 
We explain it as follows. \market\ deals with a comparatively complex setting, making it difficult for the agents to play strategically without the explicit assistance for sub-task specialization. The role-driven policy learning framework of \our\ seems to be effective here to reduce the observation-action space to explore and quickly guide them towards improved policy learning by devising opponent-aware strategic behaviors. 
\ourgame\ deals with a simpler setting, where agents learn effective strategies comparatively faster even without external asistance on implicit role division or sub-task specialization, resulting in only marginal improvement of \our\ over \bsln{.} 
Further, we observe a performance drop of \our\ agents as the training approaches for \ourgame{.} We explain this phenomenon as follows. \our\ accelerates learning strategic behavior in \ourgame, more specifically, splits the responsibility of collision and chasing the landmark among team members. However, as \bsln\ agents quickly master the strategy in the next phase of training, the advantage of \our\ almost disappear. Therefore, with the improvement of the expertise of \bsln\ team, the performance of \our\ team drops, finally resulting in marginal gain over \bsln\ agents.

\noindent \textbf{Role learning:}
Further, we investigate whether \our\ is actually helping teammates to choose diverse roles. Also, we study the strategic behaviors of \our\ agents.
In \ourgame, as we have already mentioned, though collision incurs a small penalty, one can strategically employ collision with an opponent to delay the opponent agent from reaching the landmark. 
Similarly, in \market, if the teammates choose to go for different resources, they can acquire higher rewards in future. 
To check whether algorithms are able to train their agents such strategic moves, we summarize role-discrimination statistics in fig \ref{fig:role} for both the games. 
For \ourgame\ we report the fraction of winning episodes where the agent other than the one who touches the landmark, collides with the opponents. Similarly, for \market, we report the fraction of episodes where members within the team drop different types of resources. 
In both cases, we find that \our\ agents tend to employ strategic moves more  than \bsln\ agents, which can possibly explain better performance of \our\ in direct competitions. 
The key factor behind such diverse behavior is employment of $\Lcal_D$ and $\Lcal_{Opp}$, which incentivizes the role network to assign diverse as well as opponent-aware role distributions to teammates respectively.

\noindent \textbf{Comparison with \bslncoop\ extending~\citep{wang2020roma}:} Fig \ref{fig:cooperative} compares \our\ with \bslncoop, which is a trivial extension of \roma~\citep{wang2020roma} to actor-critic framework, where we use the two regularizers used in ~\citep{wang2020roma} for role learning, but remove the opponent modeling. Instead, agents are trained using the simplified-role-module enriched \bsln\ for both of the teams separately, whereas a team it enjoys shared reward setting. Finally, agents trained under \our\ and \bslncoop\ are separately played against random \bsln\ teams selected from a common set of \bsln\ teams and average comparative results are plotted 
in fig \ref{fig:cooperative}.
The superiority of \our\ validates that our opponent-aware role-oriented framework provides an additional layer of guidance to agents in competitive team games over a straight-forward extension of \roma~\citep{wang2020roma} to actor-critics, thus emphasizing the motivation of our approach.

\noindent \textbf{Hyperparameter ($\lambda$) tuning:} Fig~[\ref{fig:vary_decay_catch_goal}-\ref{fig:vary_decay_market}] presents performance of \our\ teams against randomly chosen \bsln\ teams for different value of decay parameter $\lambda$. 
Both in \ourgame\ and \market, performance significantly varies based on chosen $\lambda$, with $\lambda=0.5$ giving the best results. $\lambda=0.9$ results in a slower decay of role loss, hence resulting in greater instability in policy learning, while $\lambda=0.1$ results in too fast a decay. This experiment confirms our intuitive hypothesis from real-world experience, that sustained role evolution beyond a point is detrimental to team performance.

%

\noindent \textbf{Comparison with ablations:} 
Fig. [\ref{fig:ablations_catch_goal}-\ref{fig:ablations_market}] compare \our\ with three ablations of \our\, which minimize either $\Lcal_D$ or $\Lcal_{MI}$ or $\Lcal_D+\Lcal_{MI}$, whereas \our\ minimizes $\Lcal_D + \Lcal_{MI}+\Lcal_{Opp}$. In both games, \our\ is observed to outperform the rest. 
In \market\, we also observe $\Lcal_{MI}$ learns roles more identifiable with trajectory and therefore training better dropping skill, in comparison with the other two baselines. 

\if{0}
For \ourgame, now we take a deeper look by observing the event of touching landmark for each agent individually in Fig. \ref{fig:tournament-landmark-agent-wise}. It is found that in case of \our\ one agent learns faster than the rest, resulting in initial advantages of team \our\ over team \bsln. But the other \our\ agent barely shows any improvement in touching the landmark. \bsln\ agents learn slower, but, the variation in fellow teammates' learning rate is smaller. \bsln\ slowly trains both of the agents, simultaneously, eventually resulting in cumulative superior performance. 
This observation indicates employing role network fastens the learning process and useful in shorter term.
However, if both skills are not explicitly maximizing rewards, we may find that eventually this role distinction within teammates may result in higher variance in the team members' skill-specific expertise, resulting in slowing down the overall team performance, as in \ourgame. 
On the other hand, this did not happen in \market. As a reason, we can observe that, in \market, agents within a team try to pick different resources as a result of role emergence. As both of these actions implicitly increase the rewards, learning specific roles does not hamper retaining the higher rewards throughout, unlike \ourgame. 
\fi

%% file: 300conclusion.tex
\vspace{2mm}
\section{Conclusion and Future work}
\vspace{2mm}
We propose an algorithm for opponent-aware role-based learning in actor-critic framework targeted towards team competitions.  Our algorithm combines a self-role encoder and an opponent-role predictor in actor-critic framework for learning an optimal policy.  We analyze our approach in two scenarios where we show how our method improves the quality of learning. As a future direction, we intend to extend our algorithm to more general settings where multiple teams can compete among themselves.